\documentclass[manuscript]{aastex}

\renewcommand{\vec}[1]{\mbox{\boldmath $#1$}}
\newcommand{\mrm}[1]{\mathrm{#1}}
\slugcomment{Not to appear in Nonlearned J., 45.}

\shorttitle{Superslow cross-field propagation}
\shortauthors{Kaneko et al.}

\begin{document}


\title{Apparent cross-field superslow propagation of magnetohydrodynamic waves in solar plasmas}


\author{T. Kaneko\altaffilmark{1}, M. Goossens\altaffilmark{2}, R. Soler\altaffilmark{3},  J. Terradas\altaffilmark{3}, T. Van Doorsselaere\altaffilmark{2}, T. Yokoyama\altaffilmark{1}, 
and A. N. Wright\altaffilmark{4}.}
\email{kaneko@eps.s.u-tokyo.ac.jp}

\altaffiltext{1}{Department of Earth and Planetary Science, The University of Tokyo,
  7-3-1 Hongo, Bunkyo-ku, Tokyo, 113-0033, Japan}
\altaffiltext{2}{Centre for Mathematical Plasma Astrophysics, Katholieke Universiteit Leuven, 
Celestijnenlaan 200B,bus 2400, B-3001 Herverlee, Belgium}
\altaffiltext{3}{Departament de F\'{i}sica, Universitat de les Illes Balears, E-07122 Palma de Mallorca, Spain}
\altaffiltext{4}{School of Mathematics and Statistics, University of St Andrews, St Andrews, KY16 9SS, UK }


\begin{abstract}
In this paper we show that the phase mixing of continuum Alfv\'{e}n waves  and/or  continuum  slow waves in magnetic structures of the solar atmosphere as, e.g., coronal arcades, can create the illusion of wave propagation across the magnetic field. This phenomenon could be erroneously interpreted as fast magnetosonic waves. The cross-field propagation due to phase mixing of continuum waves is apparent because there is no real propagation of  energy across the magnetic surfaces.   We investigate the continuous Alfv\'{e}n and slow spectra in 2D Cartesian equilibrium models  with a purely poloidal magnetic field. We show that  apparent superslow propagation across the magnetic surfaces in solar coronal structures is a consequence of the existence of continuum  Alfv\'{e}n waves and continuum slow waves that naturally live on those structures and phase mix as time evolves. The apparent cross-field phase velocity is related to the  spatial variation of the local Alfv\'{e}n/slow frequency across the magnetic surfaces and is  slower than the Alfv\'{e}n/sound velocities for typical coronal conditions. Understanding the nature of the apparent cross-field  propagation  is important for the correct analysis of  numerical simulations and the correct interpretation of observations.
\end{abstract}


\section{Introduction}

\begin{figure}
  \begin{center}
    \includegraphics[bb=0 0 708 1133,scale=0.45]{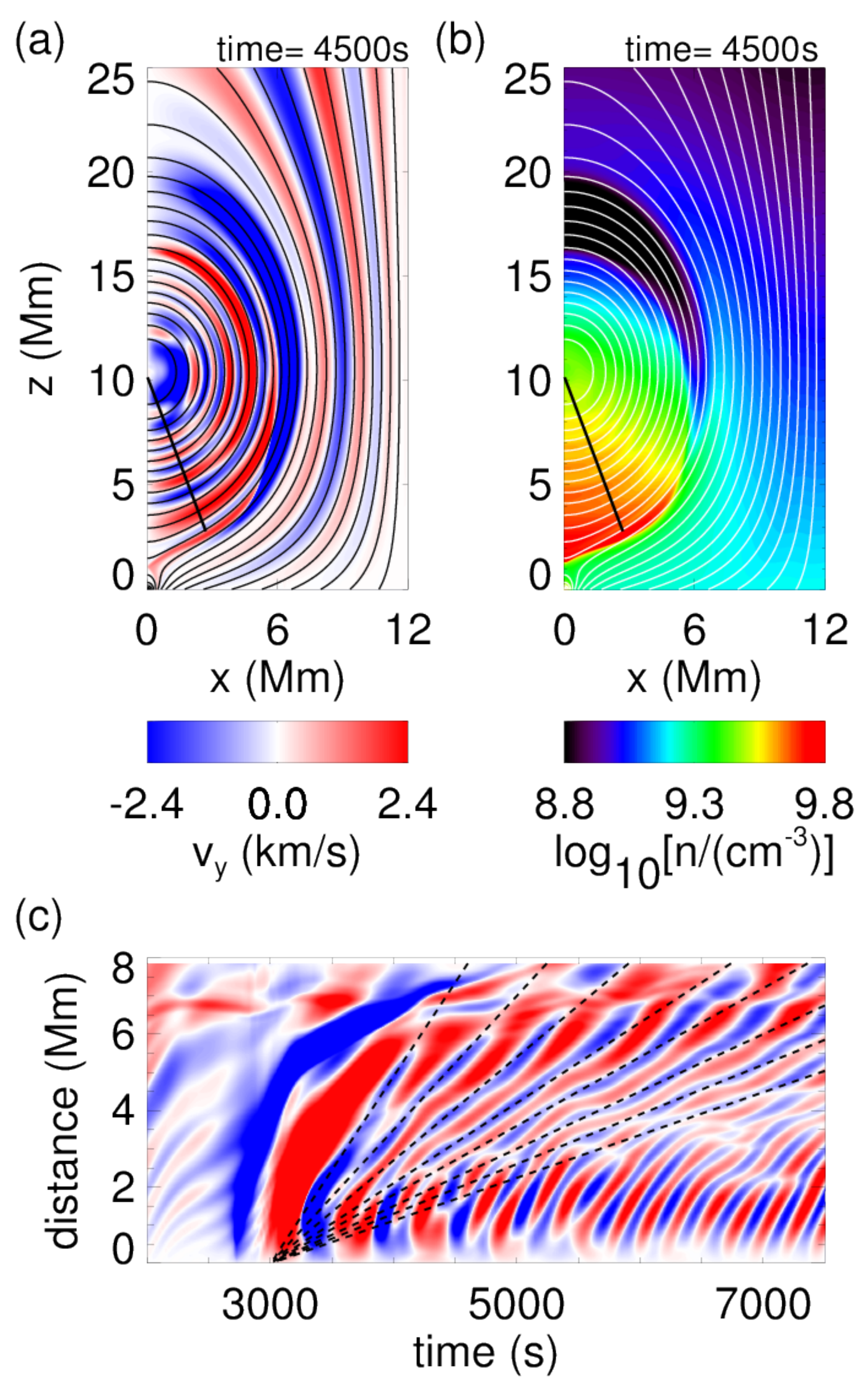}
    \caption{Superslow propagation in \citet{KanekoYokoyama2015}. 
      Panels (a) and (b) show the velocity component perpendicular to the plane 
      and number density at a certain time, respectively. 
      The thin solid lines represent magnetic field. The thick solid line is the slit. 
      Panel (c) shows the time evolution of velocity component perpendicular to the plane along the slit in panel (a). 
      The horizontal and vertical axes represent time and distance from the center of the flux rope, respectively.}
    \label{snapshots_los} 
 \end{center}
\end{figure}

Recent  numerical simulations of magnetohydrodynamic (MHD) waves in coronal arcades \citep{Rial2010ApJ,Rial2013ApJ}  and in the interior of prominences \citep[][shown later]{KanekoYokoyama2015} have revealed the presence of MHD waves propagating  across the magnetic surfaces at slow velocities. 
It is standard to associate propagation across magnetic surfaces with fast magnetosonic MHD waves. However, the interpretation in terms of fast magnetosonic waves poses a problem  since the apparent velocity of the cross-field propagation reported in those numerical studies is  slower than that associated with a fast MHD wave and even a slow MHD wave. Here we show an example of cross-field superslow propagation. 
Figure. \ref{snapshots_los} (a) and (b) show snapshots at a certain time of the simulation in 
\citet{KanekoYokoyama2015}, and Fig. \ref{snapshots_los} (c) shows the time evolution of  
the velocity component perpendicular to the plane along the slit in panel (a).  
In this simulation, 
radiative condensation happens at around time of 3000 s, 
and the waves are excited inside the flux rope.
In Fig. \ref{snapshots_los} (c), 
at the region apart from the center of the flux rope (distance of 2--7 Mm) 
we clearly find waves which propagate outward and whose propagation speeds are decreasing with time.
The propagation speeds are 1--5 km/s (as shown by dashed lines in panel (c)),  
much slower than the characteristic propagation speeds of the fast mode 
($\sim $ 160 km/s in our simulation settings) and even the slow mode ($\sim $ 70 km/s). 
We think that the  superslow propagation is explained as the apparent effect caused 
by phase mixing of standing Alfv\'{e}n or slow waves trapped in the closed loops of the flux rope.  
In the present paper, as a first step, it is argued that in magnetic structures of the solar corona as, e.g., magnetic arcades, the phase mixing of continuum Alfv\'{e}n waves  and/or  continuum  slow waves can create the illusion of MHD waves propagating across magnetic surfaces at velocities smaller than the characteristic sound and Alfv\'{e}n  velocities of the plasma. This cross-field propagation is apparent because there is no real propagation of wave energy across the magnetic field.

Continuum  Alfv\'{e}n waves and continuum  slow waves  live on individual magnetic surfaces and are associated with the Alfv\'{e}n continuum and slow continuum of the linear MHD spectrum \citep{Appert1974PhFl}. Each magnetic surface can oscillate at its own local Alfv\'{e}n frequency and local slow frequency without interaction with neighbouring magnetic surfaces in ideal MHD and with negligible interaction in non-ideal MHD. If the continuum Alfv\'{e}n/slow waves on a collection of neighbouring magnetic surfaces are excited  each at their own local Alfv\'{e}n/slow frequencies, an observer would see an apparent phase propagation across the magnetic surfaces due to the  variation of the local Alfv\'{e}n/slow frequency across those surfaces \citep{Rial2010ApJ,Rial2013ApJ,KanekoYokoyama2015}. The apparent phase velocity  is related to the  spatial variation of the local Alfv\'{e}n/slow frequency across the magnetic surfaces and is  slower than the Alfv\'{e}n/sound velocities for typical coronal conditions. The apparent propagation may be  misleading for the analysis of simulations and observations, since this phenomenon could naturally be interpreted as fast MHD waves.  Therefore, understanding the nature of the apparent wave propagation is important for the correct analysis of  numerical simulations and the correct interpretation of observations.

Computations of the continuous spectrum that are relevant for the present investigation can be found in, e.g., \citet{PoedtsGoossens1987SoPh,PoedtsGoossens1988A&A,PoedtsGoossens1991SoPh}, \citet{Oliver1993A&A}, \citet{TirryPoedts1998A&A}, \citet{Arregui2004A&A,Arregui2004ApJ} and \citet{Terradas2013ApJ}.  These investigations are concerned with 2D equilibrium models in Cartesian geometry that are invariant in the perpendicular direction to the 2D plane ($y$-direction). 
\citet{PoedtsGoossens1987SoPh,PoedtsGoossens1988A&A,PoedtsGoossens1991SoPh} computed the continuous spectrum of ideal MHD waves in 2D solar coronal loops and arcades. They dealt with equilibrium models with a purely poloidal magnetic field \citep{PoedtsGoossens1987SoPh,PoedtsGoossens1988A&A} and a mixed poloidal and toroidal magnetic field \citep{PoedtsGoossens1991SoPh}. They explicitly determined how the slow continuum frequencies and the Alfv\'{e}n continuum frequencies change across the magnetic surfaces for specific choices of the magnetic field and equilibrium density. \citet{Oliver1993A&A} computed the Alfv\'{e}n continuous spectrum of a pressureless coronal arcade with a poloidal potential magnetic field. They neglected gravity and they removed the slow part of the spectrum by using the  assumption that the plasma is pressureless.  \citet{TirryPoedts1998A&A} studied MHD waves in potential arcades as \citet{Oliver1993A&A}. They determined the variation of the frequencies of Alfv\'{e}n continuum modes across the magnetic surfaces for a specific density profile. Subsequently \citet{TirryPoedts1998A&A} studied the coupling of Alfv\'{e}n continuum modes and fast modes in the resistive driven problem for $k_y \neq 0$, where $k_y$ denotes the wavenumber in the $y$-direction (the direction in the magnetic surfaces perpendicular to the magnetic field lines). \citet{Arregui2004A&A,Arregui2004ApJ} studied MHD waves in potential arcades as \citet{Oliver1993A&A} and in force free arcades. They determined the variation of the frequencies of  Alfv\'{e}n  continuum modes across the magnetic surfaces for a specific density profile corresponding to $\delta = 6$ in the notation of \citet{Oliver1993A&A}.  They used their results on Alfv\'{e}n continuum modes for a purely poloidal field and $k_y = 0$  as starting point to understand the coupling of  Alfv\'{e}n continuum modes and fast waves in more complicated cases.  \citet{Terradas2013ApJ} computed the slow and Alfv\'{e}n continuum for a 2D prominence model with a purely poloidal magnetic field and gravity.

The aim of the present paper is to show that  apparent superslow propagation across the magnetic surfaces in solar coronal structures is a consequence of the existence of continuum  Alfv\'{e}n waves and continuum slow waves. To this end we investigate the continuous spectrum for 2D equilibrium models in Cartesian geometry that are invariant in the $y$-direction and have a purely poloidal magnetic field. The actual equilibrium configurations that we have in mind are 2D coronal arcades \citep[e.g.][]{Oliver1993A&A}. The assumption that there is no toroidal magnetic field leads to two separate continuous parts. It simplifies the mathematical analysis and enables us to understand the essential mechanism behind the apparent superslow propagation.

The plan of the paper is as follows. In Section 2 we  review the concept of continuous spectrum of linear ideal MHD and  we recall the equations that govern the continuous spectrum for a 2D magnetostatic equilibrium in Cartesian coordinates with a purely poloidal magnetic field.  In Section 3 we discuss the solutions for the Alfv\'{e}n continuum waves and slow continuum waves. The apparent cross-field propagation caused by the phase mixing of continuum  waves is studied in Section 4.  In Section 5 we use the theory of apparent superslow propagation due to continuum Alfv\'{e}n waves to explain the superslow propagation observed in the numerical simulations by \citet{KanekoYokoyama2015}. Conclusions are formulated in Section 6.

\section{The continuous spectrum}

The continuous part of the linear spectrum of ideal MHD was first studied for 1D magnetostatic equilibrium models. 
\cite{Appert1974PhFl} were the first to give a rigorous proof that  the linear spectrum of ideal MHD contains 
a continuous part. Their analysis applied to a 1D axisymmetric circular plasma cylinder, 
known in the plasma physics literature as the diffuse linear pinch. 
Waves belonging to the continuous part of the spectrum are recognized by their singular behaviour 
at a magnetic surface. In the case of a 1D magnetostatic equilibrium model (e.g. the plasma slab, 
the diffuse linear pinch) the linear MHD equations can be reduced to the classic Hain-Lust equation.  
The values of $\sigma^2$ that  correspond to the mobile regular singular points of the Hain-Lust equation 
\citep{hain1958normal,Goedbloed1972PhFl}  are associated   with non-square integrable solutions and define two separate continuous parts of the spectrum, namely the Alfv\'{e}n continuum and the cusp or slow continuum 
\citep[see e.g.][]{goedbloed1983lecture,GoedbloedPoedts2004prma.book,Goossens1991,Sakurai1991,Goossens1992}.   
The solutions that correspond to the Alfv\'{e}n continuum and slow continuum are localized on  the magnetic surfaces where the resonant conditions of the respective wave dispersion relations are satisfied.  In addition they are characterized by motions in the magnetic surfaces respectively perpendicular and parallel to the magnetic field lines.

For 1D equilibrium models the determination of the frequencies of the continuous part of spectrum is relatively straightforward: put the coefficient function of the highest derivative in the Hain-Lust  equation equal to zero. The resonant frequencies are given by simple algebraic relations. For 2D equilibrium models matters are more complicated. The equations for the linear motions are partial differential equations. The continuous spectrum is redefined as the collection of frequencies for which the solutions show non-square integrable singularities at a flux surface $\Psi = \Psi_0$. \citet{Pao1975} and \citet{Goedbloed1975} were the first to determine independently the equations that govern the continuous part of the linear ideal spectrum for 2D toroidal equilibrium configurations in the context of fusion plasma physics. They also derived basic properties of the continuous spectrum that  do not depend on the details of the magnetic field. In particular they showed that in the general case of a mixed poloidal and toroidal magnetic field the Alfv\'{e}n continuum and the cusp continuum become coupled and the continuum modes are no longer polarized purely parallel and purely perpendicular to the magnetic field lines. When the magnetic field is purely poloidal the Alfv\'{e}n continuum and the slow continuum remain uncoupled and the continuum solutions are polarized  as in  the 1D case of the diffuse linear pinch.

In the astrophysical context, \citet{Poedts1985} and \citet{Goossens1985} derived  the equations that govern the  continuous spectrum  for 2D equilibrium models in the presence of gravity.  \citet{Poedts1985} considered a toroidal equilibrium model in cylindrical coordinates with invariance in the $\varphi$-direction.  \citet{Goossens1985} used a Cartesian model with invariance in the $y$-direction. \citet{Poedts1985} and \citet{Goossens1985} confirmed the result known in fusion plasma physics that the two continua are coupled when the magnetic field has  a component in the ignorable direction respectively $B_{\varphi}$ and $B_y$. In that situation both continua are affected by gravity. For a purely poloidal magnetic field the two continua are uncoupled and the corresponding solutions have the  classic properties known from the analysis of the diffuse linear pinch. Here the Alfv\'{e}n continuum is not affected by gravity, but  the slow continuum is affected  and it might be better referred  to as the slow-gravity continuum. The singular solutions of the continuum Alfv\'{e}n waves for 2D magnetostatic equilibrium models 
with a purely poloidal magnetic field were discussed  in detail by \citet{ThompsonRight1993JGR}, \citet{WrightThompson1994} 
and \citet{TirryGoossens1995}. 

\subsection{Continuous spectrum for a 2D equilibrium}

In the present investigation we use the equations for the continuous part of the linear spectrum formulated by \citet{Goossens1985}. These authors derived the equations that govern the continuous part of  linear ideal MHD for  2D equilibrium configurations in Cartesian geometry that are  invariant in the $y$-direction. The basic equations for the magnetostatic equilibrium and the linear motions superimposed on this equilibrium can be found in Section 2 of \citet{Goossens1985}. We recall the necessary equations from \citet{Goossens1985} and add new information. The equilibrium quantities are functions of the Cartesian coordinates $x$ and $z$ but not of $y$.  \citet{Goossens1985} implicitly specified the dependence on the ignorable coordinate $y$ and time $t$ as 
\begin{equation}
\exp(i k_y y - i \sigma t)
\label{yt}
\end{equation}
with $k_y$ the wave number in the $y$-direction and $\sigma$ the frequency. It is standard practice to split  the equilibrium magnetic field  in a poloidal magnetic field $\vec{B}_p$ and a toroidal magnetic field $\vec{B}_t = B_y \vec{1}_y$.  In the present paper we deal  with  equilibrium configurations with a purely poloidal magnetic field.  In what follows $B_y = 0$.

 The poloidal magnetic field is written in terms of a magnetic flux function $\Psi(x,z)$ as
\begin{equation}
\vec{B}(x,z) = \vec{B}_p(x,z) = - \nabla \Psi(x,z) \times \vec{1}_y = 
\frac{\displaystyle \partial \Psi}{\displaystyle  \partial z} \vec{1}_x -  \frac{\displaystyle \partial \Psi}{ \displaystyle \partial x} \vec{1}_z
\label{B1}
\end{equation}
where $\vec{1}_x, \;\vec{1}_y, \; \vec{1}_z$  are the unit vectors in the $x$-, $y$- and $z$-directions. The definition of $\vec{B}_p$ with the use of the flux function $\Psi$ implies  that $ \vec{B}_p \cdot \nabla \Psi = 0$.
\citet{Goossens1985}  used  a local system of flux coordinates $(\Psi, y, \chi) $ with $\chi$ the poloidal variable. All equilibrium variables are functions of $\Psi$ and $\chi$ but not of $y$. The equilibrium magnetic field has components $(0, 0, B_{\chi})$ in the $(\Psi, y, \chi)$ system of coordinates. Expressions for the operators $\nabla, \nabla^2, \mbox{div}, \mbox{rot}$  can be found in Equations (7) - (10) of \citet{Goossens1985}. The unit vector normal to the flux surfaces is $\vec{1}_{\Psi}$ and the unit vector in the magnetic surfaces parallel to the poloidal magnetic field is $\vec{1}_{\chi}$.  For completeness, we note that $\vec{1}_y$  is the unit vector in the magnetic surfaces perpendicular to the poloidal magnetic field lines and $k_{y}$ is the wave number in the direction of $\vec{1}_{y}$. Hence  $\vec{1}_{\chi} = \vec{1}_{\parallel}, \;\;\vec{1}_y = \vec{1}_{\perp}$.

The  local system of flux coordinates is orthogonal so that 
$ \nabla \chi \cdot \nabla \Psi = 0$. Hence 
\begin{equation}
\nabla \chi  = \lambda(x,z)  \vec{B}_p
\label{lambda}
\end{equation}
with $\lambda(x,z)$  a function that we can choose freely. 
The Jacobian $J$ of the transformation of the Cartesian system of coordinates $(x,y,z)$  to that of the local system of orthogonal flux coordinates $(\Psi, y, \chi) $  and the elementary length in the local system of orthogonal flux coordinates are
\begin{eqnarray}
J & = & \frac{1}{\displaystyle   B_{\chi} \mid\nabla  \chi\mid } \label{J1} \\
(ds) ^2 & =& \frac{\displaystyle 1} {\displaystyle B_{\chi}^2} (d \Psi)^2 + (d y)^2 + J^2 B_{\chi}^2 (d \chi)^2
\label{ds1}
\end{eqnarray}

We use Equations (59)-(60) of \citet{Goossens1985}. They are two uncoupled ordinary differential equations for respectively $\xi_y$ and $\xi_{\chi} $  on a given magnetic surface $\Psi = \Psi_0$. The independent variable is the coordinate along the field line, $\chi$. The actual equilibrium configurations that we have in mind are 2D arcades as studied by 
\citet{PoedtsGoossens1987SoPh,PoedtsGoossens1988A&A,PoedtsGoossens1991SoPh,Oliver1993A&A,TirryPoedts1998A&A,Arregui2004ApJ,Arregui2004ApJ,Rial2010ApJ,Rial2013ApJ} and \citet{Terradas2013ApJ}.
A graphical representation of the magnetostatic configuration can be found in \citep{Oliver1993A&A} and \citet{Rial2010ApJ}.
Equations (59) - (60) of \citet{Goossens1985} are
\begin{eqnarray}
\sigma^2 \xi_y  & = &  - \frac{1}{\mu \rho_0 } (F^2 ) \; \xi_y  \label{CAW1}  \\
& & \nonumber \\
\sigma^2 \xi_{\chi}  & = & \left \{ \frac{\displaystyle v_S^2}{\displaystyle v_S^2 + v_A^2} N_{\chi}^2 
+ \frac{\displaystyle 1}{\displaystyle J B_{\chi}^2} \frac{\displaystyle \partial}{\displaystyle  \partial \chi}  \left ( \frac{\displaystyle v_S^2}{\displaystyle v_S^2 + v_A^2}  \frac{\displaystyle 1}{\displaystyle J}
\frac{\displaystyle \partial  \Phi_0}{\displaystyle \partial  \chi} 
\right )  \right \}\xi_{\chi}, \nonumber \\
& -  & \frac{1}{\displaystyle \rho_0 B_{\chi}}
F \left \{ \rho_0 \; v_C^2  \;
F \left ( \frac{\displaystyle \xi_{\chi}} {\displaystyle B_{\chi}} \right ) \right \}, 
\label{CSW1}
\end{eqnarray}
The operator $F$ is given by 
\begin{equation}
F= \frac{\displaystyle 1}{ \displaystyle J} \frac{\displaystyle \partial }{\displaystyle \partial \chi}.
\label{Fster1}
\end{equation}
Note that \citet{Goossens1985} used  the notation $F^{\star}$ in stead of $F$. In these equations, $\rho_0, p_0, \Phi_0$ are the equilibrium density, pressure, and gravitational potential. In turn, $v_S^2,$ $v_A^2$, and $v_C^2$  are the square of the local speed of sound,  the local Alfv\'{e}n velocity,  and the local cusp (or tube) speed defined as
\begin{equation}
v_S^2   =  \frac{\displaystyle \gamma p_0}{\displaystyle \rho_0} , \;\;
v_A^2 = \frac{\displaystyle B^2}{ \displaystyle \mu \rho_0}, \;\; v_C^2 = \frac{\displaystyle v_A^2 \; v_S^2}{\displaystyle v_A^2 + v_S^2},
\label{vSvA1vC} \\
\end{equation}
where $\gamma$ is the adiabatic index and $\mu$ is the magnetic permeability. 
$ N_{\chi}^2$ is the square of the Brunt-Vais\"{a}l\"{a} frequency along the magnetic field lines. It is defined as
\begin{equation}
N_{\chi}^2   =  - \frac{\displaystyle 1} {\displaystyle J B_{\chi}}  \frac{ \displaystyle \partial \Phi_0}{ \displaystyle \partial \chi}  
\left \{ \frac{\displaystyle 1} {\displaystyle J B_{\chi}} \frac{\displaystyle 1}{\displaystyle \rho_0} \frac{\displaystyle \partial \rho_0}{\partial \chi} 
- \frac{\displaystyle 1}{\displaystyle \gamma p_0 J B_{\chi}} \frac{\displaystyle \partial p_0}{\partial \chi} \right \}.
\label{BVF1}
\end{equation}

When equations (\ref{CAW1}) and  (\ref{CSW1})   are supplemented with boundary conditions they define two uncoupled  eigenvalue problems for the frequency, $\sigma$.  When the magnetic surface is varied, the corresponding frequencies define respectively the Alfv\'{e}n continuum and the cusp or slow continuum. Note that $\xi_y = \xi_{\perp}$  and $\xi_{\chi} = \xi_{\parallel}$.
The equation for $\xi_y$ is independent of gravity and hence the Alfv\'{e}n continuum is unaffected by gravity. The coefficient function of $\xi_{\chi}$ in the equation for $\xi_{\chi}$ clearly depends on  gravity.   Hence the slow continuum is affected by gravity. Note also that the wavenumber $k_y$ does not appear in Equations (\ref{CAW1}) and  (\ref{CSW1}), so that the two continua are independent of $k_y$. The corresponding solutions spatially depend on $y$ through the factor $\exp(i k_y y)$.

\subsection{Normalized variables}

All the equations so far have been written  in terms of dimensional variables. From here on we use normalized or dimensionless variables. We introduce a reference length $L_R$ and use it to define the normalized coordinates $x_{\star}, y_{\star}, z_{\star}$,  the normalized arc length  $s_{\star}$ and the normalized components of the Lagrangian displacement $\xi_{y,\star} \; \xi_{\chi, \star}$ as 
\begin{equation}
[x,y,z]^t = L_R \; [x_{\star}, y_{\star}, z_{\star}]^t, \; \;\;\;s = L_R\; s_{\star}, \;\;\;[\xi_y, \xi_{\chi}]^t = L_R \; [\xi_{y \star}, \; \;
\;   \xi_{\chi \star}]^t
\label{xyzs-star}
\end{equation}
The operator $\nabla$ is transformed as $\nabla = (1/L_R) \;\nabla_{\star}$. Next we introduce the reference value $\Psi_R$ to normalize the magnetic flux function $\Psi$,  the variable $\chi$ and the poloidal magnetic field $\vec{B}_p$ as
\begin{eqnarray}
\Psi & =& \Psi_R \;\Psi_{\star}(x_{\star},\; y_{\star}, \;z_{\star}), \; \chi = \Psi_R \; \chi_{\star}(x_{\star},\; y_{\star}, \;z_{\star}), 
\nonumber \\
\vec{B}_p & = & B_R \;\vec{B}_{p\star}, \;\; B_R = \Psi_{R}/L_R .
\label{PsichB-star}
\end{eqnarray}
Equation (\ref{lambda}) is then
\begin{equation}
\nabla_{\star} \chi_{\star}  = \lambda(x,z) \; \vec{B}_{p \star}
\label{lambda2}
\end{equation}
A convenient  choice for the multiplicative function $\lambda(x,z)$ in (\ref{lambda2}) is 
$\lambda(x,z) = 1/(\mid B_{\chi \star}(x,z) \mid )$ 
so that
\begin{equation}
\nabla_{\star}\;\chi_{\star}= \vec{1}_{B} = \vec{1}_{\chi}, \; \mid \nabla_{\star}\chi _{\star}\mid = 1, \; J = (1/B_R^2) J_{\star}, \; 
J _{\star}= \frac{1}{B_{\chi \star}}.
\label{chi-2}
\end{equation}
The expression (\ref{ds1}) for the elementary length is in dimensionless variables 
\begin{equation}
(ds_{\star}) ^2 = \frac{\displaystyle 1} {\displaystyle B_{\chi  \star}^2} (d \Psi_{\star})^2 + (d y_{\star})^2 + (d \chi_{\star})^2
\label{ds2}
\end{equation}
Equation (\ref{ds2}) tells us that  for constant $\Psi_{\star}$  and constant $y_{\star}$ 
\begin{equation}
(ds_{\star}) ^2 = (d \chi_{\star})^2.
\label{ds3}
\end{equation}
Hence $ \chi _{\star}= s_{\star}$ with $s_{\star}$ the normalized arc length along a poloidal field line. In addition (\ref{Fster1}) 
can be simplified to
\begin{equation}
F = \frac{B_R}{L_R} \; F_{\star}, \; \;\; F_{\star} 
= B_{\chi  \star} \frac{\displaystyle \partial }{\displaystyle \partial \chi_{\star}}
= B_{\star} \frac{\displaystyle \partial }{\displaystyle \partial s_{\star}}.
\label{Fster2}
\end{equation}
The equilibrium density and equilibrium pressure $\rho_0, \;p_0\;$ are normalized by the use of the reference values $\rho_{ R}$ and 
$p_{ R}$:
\begin{equation}
\rho_0 = \rho_{ R} \;\rho_{\star}, \;\;p_0 = p_{R} \;p_{\star}
\label{p0rho0}
\end{equation}
The local Alfv\'{e}n velocity $v_A$, the local speed of sound $v_S$  and  the local cusp speed $v_C$ are 
normalized with the reference value for the local Alfv\'{e}n speed $v_{AR}$
\begin{equation}
v^2_{A R} = B_R^2 /(\mu \rho_R), \;( v_A , v_S, v_C) = v_{A R} \; (v_{A \star}, \;v_{S \star}, \;v_{C \star})
\label{vAvSvC}
\end{equation}
Expressions for $v_{A \star}, \;v_{S \star}, \;v_{C \star}$ are
\begin{equation}
v^2_{A \star} = \frac{B_{\star}^2}{\rho_{\star}}, \;\; v^2_{S \star} = \frac{\beta_R}{2} \frac{\gamma p_{\star}}{\rho_{\star}}, \;\; 
\beta_R = \frac{p_R}{B^2_R/(2 \mu)}, \;\; v_{C \star}^2 = \frac{\displaystyle v_{A \star}^2 \; v_{S \star}^2}
{\displaystyle v_{A \star}^2 + v_{S \star}^2}
\end{equation}
The equilibrium potential $\Phi_0$ is normalized as
\begin{equation}
\Phi_0 = v^2_{A R} \;\Phi_{\star}
\label{Phi0}
\end{equation}
Finally all frequencies are normalized by use of the reference Alfv\'{e}n frequency $\sigma_{A R}$ as 
\begin{equation}
\sigma^2_{A R} = \frac{v_{A R}^2}{{L_R}^{2}}, \; \; \; \sigma^2 = \sigma^2_{A R} \sigma_{\star}^2, \;\; \; N_{\chi}^2 = \sigma^2_{A R} 
N_{\chi \star}^2
\end{equation}
The dimensionless square of the Brunt-Vais\"{a}l\"{a} frequency $N_{\chi  \star}^2$ is 
\begin{equation}
N_{\chi  \star}^2  =  - g_{\chi \star} 
\left \{  \frac{\displaystyle 1}{\displaystyle \rho_{\star}} \frac{\displaystyle \partial \rho_{\star}}{\partial s_{\star}} + 
\frac{\displaystyle g_{\chi,\star}  }{\displaystyle v_{S'\star}^2} \right \}
\label{BVF2}
\end{equation}
$ g_{\chi \star}$ is the dimensionless component of gravity along the field line 
\begin{equation}
 g_{\chi \star}   = \frac{\displaystyle \partial \Phi_{\star}}{\displaystyle \partial s_{\star}} , \label{gs} 
\end{equation}
From here on only normalized quantities will be used and there is no room for confusion. Hence  we drop the subscript $\star$ for the sake of simplicity.  Equations (\ref{CAW1}) and  (\ref{CSW1})   for the Alfv\'{e}n continuum waves  and the slow continuum waves can then be rewritten as 
\begin{eqnarray}
\sigma^2 \xi_y  & = &  - \frac{1} {\rho } (B \frac{\displaystyle \partial }{\displaystyle \partial s} )
\left \{B \frac{\displaystyle \partial \xi_y}{\displaystyle \partial s} \right \}, \label{CAW2} \\
& & \nonumber \\
\sigma^2 \xi_{\chi}  & = & \left \{ \frac{\displaystyle v_S^2}{\displaystyle v_S^2 + v_A^2} \; N_{\chi}^2 
+ \frac{\displaystyle 1}{\displaystyle B}
\frac{\displaystyle \partial}{\displaystyle  \partial s}  \left ( \frac{\displaystyle v_S^2}{\displaystyle v_S^2 + v_A^2} B \;g_{\chi}
\right )  \right \}\xi_{\chi}, \nonumber \\
& -  & \frac{1}{\displaystyle \rho }\frac{\partial}{\partial s}
 \left \{\rho \; v_C^2 \; B \frac{\partial}{\partial s}
\left ( \frac{\displaystyle \xi_{\chi}} {\displaystyle B} \right ) \right \}.
\label{CSW2}
\end{eqnarray}

\section{Continuum waves}

\subsection{Alfv\'{e}n waves}\label{sec:alfven}

The  Alfv\'{e}n continuum waves are governed by equation (\ref{CAW2}).  It is an  ordinary  differential equation of second order for 
$\xi_y = \xi_{\perp}$.   We have deliberately kept the notation with the partial derivative  $\partial / \partial s$ to make it clear that we are  on a given magnetic surface $\Psi = \Psi_0$. The continuum Alfv\'{e}n waves live on individual magnetic surfaces and the motions are in the $y$-direction i.e. in the magnetic surfaces and perpendicular to the magnetic field lines. Let us rewrite
equation (\ref {CAW2}) as 
\begin{equation}
\frac{\displaystyle \partial^2 \xi_y ^2}  {\displaystyle \partial s^2} + \frac{\displaystyle 1}{\displaystyle B} \frac{\displaystyle \partial B}
{\displaystyle \partial s} \; \frac{\displaystyle \partial \xi_y }  {\displaystyle \partial s}  + \frac{\displaystyle \sigma^2}
{\displaystyle v_A^2}   \; \xi_y= 0.
\label{CAW3}
\end{equation}
Equation (\ref{CAW3}) agrees with Equation (15) of \citet{Terradas2013ApJ}, which was obtained from Equation (59) of \citet{Goossens1985}. Equation (\ref{CAW3})  is defined on the field line $\Psi = \Psi_0$. Let us denote the length of the field line  as $L(\Psi_0)$  and impose the  boundary conditions that the magnetic field lines of the arcade are anchored in the dense plasma of photosphere
\begin{equation}
\xi_{y, \Psi_0}( s=0) = \xi_{y, \Psi_0}( s=L(\Psi_0)) =0.
\label{BC-AW}
\end{equation}
The boundary conditions (\ref{BC-AW}) were also used by \citet{PoedtsGoossens1987SoPh,PoedtsGoossens1988A&A,PoedtsGoossens1991SoPh,Oliver1993A&A,Arregui2003A&A,Arregui2004A&A,Arregui2004ApJ} and \citet{Terradas2013ApJ}. Equation (\ref{CAW3}) and boundary conditions (\ref{BC-AW}) define an eigenvalue problem with eigenvalue $\sigma^2$ and eigenfunction $\xi_{y, \Psi_0}(s)$. There are infinitely many eigensolutions 
\begin{equation}
\sigma^2_{A,n}(\Psi_0), \;\; \xi_{y,A,n,\Psi_0}(s).
\label{CAW4}
\end{equation}
The notation in (\ref{CAW4}) is as follows: the subscript $A$ refers to Alfv\'{e}n waves, $\Psi_0$ refers to the fact that we are on the magnetic surface $ \Psi = \Psi_0$. The  number $n$ is related  to  the number of internal nodes of $\xi_y$ as function of $s$, i.e. along the field line. So, $n=1$ corresponds to the fundamental mode, with no internal nodes,  $n=2$  to the first overtone,  with one internal node, etc \citep[see][]{PoedtsGoossens1987SoPh,PoedtsGoossens1988A&A,Arregui2003A&A,Arregui2004A&A,Arregui2004ApJ}. When we change $\Psi_0$ from $\Psi_B$ to $\Psi_E$ each of the frequencies 
$\sigma^2_{A,n}(\Psi_0)$  maps out a continuous range of Alfv\'{e}n frequencies. Hence we have infinitely many Alfv\'{e}n continua
\begin{equation}
\sigma^2_{A,n}(\Psi), \;\; \Psi_B \leq \Psi \leq \Psi_E
\label{CAW5}
\end{equation}
where we have used $\Psi$  in stead of $\Psi_0$.

Let us turn back to equation (\ref{CAW4}). In general it does not admit closed analytical solution. The reason is that the coefficient function of the first order derivative of $\xi_y$  in the left hand member of equation (\ref{CAW4}) is in general non-zero and the coefficient function of 
$\xi_y$  in the left hand member of equation (\ref{CAW4}) is in general non-constant. The coefficient function of the first order derivative of $\xi_y$  in the left hand member of equation (\ref{CAW4}) is zero only when the magnetic field strength does not vary along the field line, i.e. when $B$ is a flux function $B = B(\Psi)$. However, there are situations where  the magnetic field strength does  vary along the field line as e.g. in \citet{PoedtsGoossens1987SoPh,PoedtsGoossens1988A&A,PoedtsGoossens1991SoPh,Oliver1993A&A,Arregui2003A&A,Arregui2004A&A,Arregui2004ApJ} and \citet{Terradas2013ApJ}. The coefficient function of  $\xi_y$  in the left hand member of equation (\ref{CAW4}) is constant if the Alfv\'{e}n velocity $v_A$ is constant i.e. when $v_A$ is a flux function $v_A = v_A(\Psi)$. The combination of  $B$ is a flux function $B = B(\Psi)$ and $v_A$ is a flux function $v_A = v_A(\Psi)$ implies that density $\rho$  is a flux function $\rho = \rho(\Psi)$. Again in general the equilibrium density is not a flux function as f.e. \citet{PoedtsGoossens1987SoPh,PoedtsGoossens1988A&A,PoedtsGoossens1991SoPh,Oliver1993A&A} and \citet{Terradas2013ApJ}. Hence,  in general equation (\ref{CAW4})  must be numerically solved. There are of course exceptions.  

\citet{Oliver1993A&A} considered a coronal arcade model and were able to obtain closed analytical solutions for the eigenfrequencies and eigensolutions of continuum Alfv\'{e}n waves by the use of a clever choice of a non-constant equilibrium magnetic field and a non-constant equilibrium density. As a  means for comparison, let us  consider the case that both the magnetic field strength $B$ and the equilibrium density $\rho$ are flux functions: $B = B(\Psi), \; \rho = \rho(\Psi)$. The equation (\ref{CAW3}) for continuum Alfv\'{e}n waves can be simplified to 
\begin{equation}
\frac{\displaystyle \partial^2 \xi_y }  {\displaystyle \partial s^2}  + \frac{\displaystyle \sigma^2}
{\displaystyle v_A^2}   \; \xi_y= 0
\label{CAW6}
\end{equation}
where now $\sigma^2 / v_A^2$ is a flux function and  independent of $s$. The solutions to equation (\ref{CAW6}) and boundary conditions 
(\ref{BC-AW}) are 
\begin{eqnarray}
\sigma^2_{A,n}(\Psi_0) & = & \frac{\displaystyle n^2 \pi^2 }{\displaystyle L^2(\Psi_0)}  v_A^2(\Psi_0) \label{CAWFR1} \\
 \xi_{y,A,n,\Psi_0}(s) & = & \delta(\Psi - \Psi_0) \sin\left(\frac{\displaystyle n \pi}{\displaystyle L(\Psi_0)} \;s \right)
\label{CAWEF1}
\end{eqnarray}
The continuum Alfv\'{e}n frequencies are
\begin{eqnarray}
 \sigma_{A,n}(\Psi) & = & \frac{\displaystyle n \pi } {\displaystyle L(\Psi)} v_A(\Psi) \nonumber \\
 & = & k_{\chi} (\Psi) \; v_A(\Psi)
\label{CAWFR2}
\end{eqnarray}
$k_{\chi} (\Psi)= n \pi / L(\Psi)$  is the local parallel wave number. Equation (\ref{CAWFR2}) or (\ref{CAWFR1}) defines infinitely many Alfv\'{e}n continua since $n =1, 2, 3, \ldots$:
\begin{equation}
[\;\mbox{min} \left( \frac{\displaystyle n \pi } {\displaystyle L(\Psi)} v_A(\Psi) \right)
\;\; ; \;\;\mbox{max} \left (\frac{\displaystyle n \pi } {\displaystyle L(\Psi)}  v_A(\Psi)  \right)\;]
\label{AC1}
\end{equation}
The  number of nodes  along the field line in the eigenfunction given by (\ref{CAWEF1}) is $n-1$. Hence,
$n=1$ corresponds to the fundamental mode without any  internal  nodes, $n=2$  to the first overtone with one internal node, etc. 
Equation (\ref{CAWFR2}) is very reminiscent of the classic result for 1D equilibrium models with a straight field. For instance for the diffuse linear pinch with a constant straight field the Alfv\'{e}n continuum frequencies are given by
\begin{equation}
\sigma_A(r) =  k_z v_A(r)
\label{CAWFR3}
\end{equation}
with $k_z$ the axial wavenumber or parallel wavenumber and  $v_A(r)$  the Alfv\'{e}n velocity that depends on the radial coordinate $r$. 
Equation (\ref{CAWFR2})  is formally the same as the result in equation (25) of \citet{Arregui2003A&A}. In general there are deviations from the simple result (\ref{CAWFR2}) when we consider 2D equilibrium models. These deviations are 
due to the fact that $B$ and $\rho$  are not flux functions but vary along field lines.

Numerical results for Alfv\'{e}n continuum frequencies for a magnetostatic equilibrium with a purely poloidal field are given by 
\citet{PoedtsGoossens1987SoPh,PoedtsGoossens1988A&A,Oliver1993A&A,Arregui2003A&A,Arregui2004A&A,Arregui2004ApJ}
and \citet{Terradas2013ApJ}. 
Our main interest is in the frequency of the fundamental continuum mode and its variation across magnetic surfaces for different magnetostatic equilibrium models. \citet{Oliver1993A&A} computed how the continuum Alfv\'{e}n frequency varies across the magnetic surfaces for different density variations obtained by varying their parameter $\delta$, namely the ratio of the magnetic scale height to the density scale height. The value of $\delta$ controls the variation of the local Alfv\'{e}n velocity with height. In their figure 2b \citet{Oliver1993A&A} plot the variation of the frequency of the fundamental continuum Alfv\'{e}n wave for different profiles of local Alfv\'{e}n velocity $(\delta = 1, 2, 3, 4, 6)$ as function of $x_0$. The parameter $x_0$ of  \citet{Oliver1993A&A} labels the magnetic surfaces and can be related to $\Psi$. \citet{Oliver1993A&A} found that  the variation of the  frequency of the fundamental Alfv\'{e}n wave across the magnetic surfaces depends on the value of $\delta$. For $\delta <3$ the frequency $\sigma_A$ is a strictly decreasing function of $x_0$
\begin{equation}
\forall x_0 \in [0 \; ;\;1] \;:\; \frac{d \sigma_A}{d x_0} < 0 
\label{Moncho1}
\end{equation}
However for $\delta \geq 3$ $\sigma_A$ is no longer a monotonous function of $x_0$. There is a critical point $x_C$ so that
\begin{eqnarray}
&& \exists x_C \in [0 \;;\; 1] : \frac{d \sigma_A}{d x_C} = 0  \nonumber \\
&& \forall x_0 \in [0 \; ;\;x_C[ \;:\; \frac{d \sigma_A}{d x_0} < 0 , \;\;\; 
\forall x_0 \in ]x_C\; ;\;1] \;:\; \frac{d \sigma_A}{d x_0} > 0
\label{Moncho2}
\end{eqnarray}
This behaviour of the frequency of the fundamental continuum Alfv\'{e}n mode was confirmed
by \citet{TirryPoedts1998A&A} for $\delta = 3$ and by \citet{Arregui2004A&A} for $\delta = 6$.

\subsection{Slow waves}\label{sec:slow}

The  slow continuum waves are governed by equation (\ref{CSW2}).  It is an ordinary  differential equation of second order for 
$\xi_{\chi}= \xi_{\parallel}$.  Again we have kept the notation with the partial derivative  $\partial / \partial s$ to make it clear that we are 
on a given magnetic surface $\Psi = \Psi_0$. The continuum  slow waves live on individual magnetic surfaces and the 
motions are in the $\chi$ direction i.e. in the magnetic surfaces and parallel  to the magnetic field lines. Let us rewrite rewrite equation (\ref {CSW2}) as 

\begin{equation}
\frac{\displaystyle \partial^2 \xi_{\chi} ^2}  {\displaystyle \partial s^2} + F(\Psi_0, s) 
\; \frac{\displaystyle \partial \xi_{\chi} }  {\displaystyle \partial s}  
+ G(\Psi_0, s) \; \xi_{\chi}= 0
\label{CSW3}
\end{equation}
The functions $F(\Psi_0, s) $ and  $ G(\Psi_0, s) $ are
\begin{eqnarray}
F(\Psi_0, s) & = & \frac{\displaystyle 1}{ \displaystyle \rho_0 \; v_C^2}  \frac{\displaystyle \partial (\rho_0 \; v_C^2) }{\partial s} -
\frac{\displaystyle 1}{ \displaystyle B}  \frac{\displaystyle \partial B}{\partial s}, \nonumber \\
G(\Psi_0, s) & = & \frac{\displaystyle \sigma^2} {\displaystyle v_C^2}  - \frac{\displaystyle \partial}{\displaystyle \partial s }
\left( \frac{ \displaystyle 1}{ \displaystyle B}  \frac{\displaystyle \partial B}{\partial s}  \right) - 
\frac{\displaystyle 1}{ \displaystyle \rho_0 \; v_C^2}  \frac{\displaystyle \partial (\rho_0 \; v_C^2) }{\partial s}
\frac{\displaystyle 1}{ \displaystyle B}  \frac{\displaystyle \partial B}{\partial s} \nonumber \\
& & +  \frac{\displaystyle g_{\chi} }{\displaystyle v_A^2} \;
\left \{  \frac{\displaystyle 1}{\displaystyle \rho_0} \frac{\displaystyle \partial \rho_0}{\partial s} + 
\frac{\displaystyle g_{\chi}  }{\displaystyle v_S^2} \right \}  - \frac{\displaystyle 1}{\displaystyle v_C^2} 
\frac{\displaystyle \partial}{\displaystyle \partial s } \left(\frac{\displaystyle v_C^2}{\displaystyle v_S^2} \right) g_{\chi} \nonumber \\
& & - \frac{\displaystyle 1}{\displaystyle v_S^2} \left( g_{\chi}  \frac{ \displaystyle 1}{ \displaystyle B}  \frac{\displaystyle \partial B}{\partial s} 
+ \frac{\displaystyle \partial g_{\chi}}{\partial s} \right).
\label{CSWFG}
\end{eqnarray}
Equation (\ref{CSW3}) and equation (\ref{CSWFG}) agree with equations (12) - (14) of \citet{Terradas2013ApJ} when it is taken into account that 
$g_{\chi}$ of the present paper is equal to $- g_s$ of \citet{Terradas2013ApJ}.

Equation (\ref{CSW3})  is defined on the field line $\Psi = \Psi_0$. Let us  impose the  boundary conditions 
\begin{equation}
\xi_{\chi, \Psi_0}( s=0) = \xi_{\chi, \Psi_0}( s=L(\Psi_0)) =0.
\label{BC-SW}
\end{equation}
In the same way as for the continuum Alfv\'{e}n waves Equation (\ref{CSW3}) and boundary conditions (\ref{BC-SW}) define an eigenvalue problem with eigenvalue $\sigma^2$ and  eigenfunction $\xi_{\chi, \Psi_0}(s)$. There are infinitely many eigensolutions 
\begin{equation}
\sigma^2_{S,n}(\Psi_0), \;\; \xi_{\chi,S,n,\Psi_0}(s).
\label{CSW4}
\end{equation}
The notation in equation (\ref{CSW4}) is similar to that used for continuum Alfv\'{e}n waves in equation (\ref{CAW4}).  
When we change $\Psi_0$ from $\Psi_B$ to $\Psi_E$ each of the frequecies 
$\sigma^2_{S,n}(\Psi_0)$  maps out a continuous range of Alfv\'{e}n frequecies. Hence we have infinitely many slow continua

\begin{equation}
\sigma^2_{S,n}(\Psi), \;\; \Psi_B \leq \Psi \leq \Psi_E .
\label{CSW5}
\end{equation}

Let us now turn back to equation (\ref{CSW3}). In general it does not admit closed analytical solutions. Also gravity is an ingredient that complicates simple mathematical analysis. Let us consider the case that gravity is absent. Equation (\ref{CSW2}) or (\ref{CSW3}) 
can then be simplified to 
\begin{equation}
\sigma^2 \xi_{\chi}   = -
 \frac{1}{\displaystyle \rho }\frac{\partial}{\partial s}
 \left \{\rho \; v_C^2 \; B \; \frac{\partial}{\partial s}
\left ( \frac{\displaystyle \xi_{\chi}} {\displaystyle B} \right ) \right \}.
\label{CSWNoG1}
\end{equation}
In general the equilibrium quantities $\rho, B, v_S^2, v_A^2, v_C^2 $ are not flux functions and do also depend on $s$.
As a  means for comparison, we consider the case that  the magnetic field strength $B$, the equilibrium density $\rho$  and the cusp velocity are flux functions: $B = B(\Psi), \; \rho = \rho(\Psi), \; v_C(\Psi)$. The condition that the density and the cusp velocity are flux functions implies that also pressure $p$ and temperature $T$ are flux functions : $p = p(\Psi), \; T = T(\Psi)$.  
Equation (\ref{CSWNoG1})  for continuum slow waves can then be simplified to 

\begin{equation}
\frac{\displaystyle \partial^2 \xi_{\chi} ^2}  {\displaystyle \partial s^2}  + \frac{\displaystyle \sigma^2}
{\displaystyle v_C^2}   \; \xi_{\chi}= 0
\label{CSWNoG2}
\end{equation}
where now $\sigma^2 / v_C^2$ is a flux function and  independent of $s$. 
Equation (\ref{CSWNoG2}) is identical to equation (\ref{CAW6}) for continuum Alfv\'{e}n waves with $v_A^2$ replaced with $v_C^2$.
Hence we can repeat the analysis for continuum Alfv\'{e}n waves from equation (\ref{CAWEF1}) up to (\ref{AC1}). In particular the continuum slow frequencies are 
\begin{eqnarray}
 \sigma_{S,n}(\Psi) & = & \frac{\displaystyle v_S}{\displaystyle \sqrt{v_S^2 + v_A^2}} \sigma_{A,n}(\Psi) \nonumber \\
  & = & k_{\chi} (\Psi) \; v_C(\Psi)
\label{CSWFR1}
\end{eqnarray}
The local parallel wave number $k_{\chi} (\Psi)$ is given by the same expression as for Alfv\'{e}n waves (\ref{CAWFR2}).
As for the Alfv\'{e}n continuum we can refer to the classic result for 1D equilibrium models with a straight field. For instance for the diffuse linear pinch with a constant straight field the Alfv\'{e}n continuum frequencies are given by
\begin{equation}
\sigma_S(r) =  k_z v_A(r)
\label{CSWFR2}
\end{equation}
with $k_z$ the axial wavenumber or parallel wavenumber and  $v_A(r)$  the Alfv\'{e}n velocity that depends on the radial coordinate $r$. 
In general there are deviations from the simple result (\ref{CSWFR1}) when we consider 2D equilibrium models. These deviations are 
due to the fact that $\rho, B, v_S^2, v_A^2, v_C^2 $ are not flux functions and  also  vary along the field lines and depend on $s$.

Numerical results for slow continuum frequencies for a magnetostatic equilibrium with a purely poloidal field are given by 
\citet{PoedtsGoossens1987SoPh,PoedtsGoossens1988A&A} and \citet{Terradas2013ApJ}.  
In particular \citet{PoedtsGoossens1988A&A} show that the variation of the frequency of the slow continuum modes depends on the structure of the magnetic field, the density stratification and on the plasma beta $\beta$. The variation of the frequency of the slow continuum modes across the magnetic surfaces can be both monotonic and non-monotonic with a local minimum as can be seen in figure 4 of \citet{PoedtsGoossens1988A&A}. 

\subsection{Closed magnetic surfaces}
In this subsection, we consider waves on closed magnetic flux surfaces.  
These closed surfaces could correspond to the nested flux surfaces near the core of a prominence, which are detached from the lower atmospheric layers (at least in a 2D cut of the model, see e.g. the simulations in \citet{KanekoYokoyama2014,KanekoYokoyama2015}). 
For example, we can consider a magnetic field with flux surfaces that are concentric circular cylinders. Because of the assumed $y$-invariance of equilibrium configuration and in particular of the equilibrium magnetic field, we can concentrate on the $(x,z)$ plane. The intersections of the flux surfaces with the $(x,z)$ plane define concentric circular field lines with prescribed length $L(\Psi )=2\pi R(\Psi )$ where $L(\Psi )$ and $R(\Psi )$ denote the length of the circular field and its radius on each closed magnetic surface. Note that a possible magnetic field that satisfies the condition for magneto-static equilibrium is $B(\Psi )=B_{0}(R_{0}/R(\Psi )) $. Otherwise, we need a pressure gradient in the radial direction for magneto-static equilibrium.
On these closed flux surfaces different boundary conditions have to be considered, because Eq.~(\ref{BC-AW}) and (\ref{BC-SW}) assume that the velocity perturbations are suppressed in a lower atmospheric layer with high inertia. The boundary conditions on closed magnetic surfaces are modified to 
\begin{equation}
	\xi_{y, \Psi_0}( s=0) = \xi_{y, \Psi_0}( s=L(\Psi_0)) \quad \mbox{and} \quad \xi_{\chi, \Psi_0}( s=0) = \xi_{\chi, \Psi_0}( s=L(\Psi_0)).
	\label{eq:periodicbc}
\end{equation}
The difference is that the velocity components do not need to be zero at $s=0=L(\Psi_0)$, but that the velocity components just need to be periodic functions of $s$, because the latter is a periodic coordinate as well. \par
Let us first focus on the Alfv\'en wave solutions, and let us once again consider that the magnetic field $B=B(\Psi)$ and the density $\rho=\rho(\Psi)$ are flux functions. As explained previously, the eigenvalue problem for Alfv\'en waves then reduces to Eq.~(\ref{CAW6}):
\begin{equation}
	\frac{\displaystyle \partial^2 \xi_y }  {\displaystyle \partial s^2}  + \frac{\displaystyle \sigma^2}
{\displaystyle v_A^2}   \; \xi_y= 0,
\end{equation}
with boundary condition
\begin{equation}
	\xi_{y, \Psi_0}( s=0) = \xi_{y, \Psi_0}( s=L(\Psi_0)).
\end{equation}
This is of course a well known problem, with a standard set of solutions. The general solution is
\begin{equation}
	\xi_{y, \Psi_0}=\delta(\Psi-\Psi_0)\left(\sum_{n'=1}^\infty A_{n'} \sin{\left(\frac{2\pi n'}{L(\Psi_0)} s\right)} + B_{n'} \cos{\left(\frac{2\pi n'}{L(\Psi_0)} s\right)}\right),
	\label{eq:periodiceigfct}
\end{equation}
where we have used the notation $n'$ for half the number of nodes along the flux surface (and it is thus slightly different than the meaning in Subsect.~\ref{sec:alfven} and \ref{sec:slow}). As expression for the Alfv\'en frequency continuum $\sigma_{A,n'}$ we thus find
\begin{equation}
	\sigma_{A,n'}(\Psi)=\frac{2\pi n'}{L(\Psi)}v_A(\Psi).
	\label{eq:periodiccont}
\end{equation}
\par
Analogously, one may derive the expression for the slow continuum and their eigenfunction in such a configuration of closed flux surfaces. As in Subsect.~\ref{sec:slow}, the eigenfunction will have the same form as Eq.~(\ref{eq:periodiceigfct}), but then for the $\xi_\chi$ component. Likewise, the continuum frequencies for the slow waves will be
\begin{equation}
	\sigma_{A,n'}(\Psi)=\frac{2\pi n'}{L(\Psi)}v_C(\Psi).
        \label{eq:periodiccont_s}
\end{equation}

\section{Apparent cross-field propagation due to phase mixing of continuum waves}

In this Section we show how the phase mixing of Alfv\'{e}n/slow  continuum  waves creates the illusion of wave propagation across the magnetic surfaces. We stress that this cross-field propagation is not real and derive the apparent propagation phase velocity. Since the analysis is similar for Alfv\'{e}n waves  and slow waves, first we focus on the case of Alfv\'{e}n waves and later we extend the results to slow waves.

Let us consider a situation where standing Alfv\'{e}n  continuum waves each with their own continuum frequency are excited on magnetic surfaces 
$\Psi_B \leq \Psi \leq \Psi_E$ with amplitude $A(\Psi)$   so that
\begin{equation}
\xi_y(\Psi, s,t) = A(\Psi) f_{A, \Psi}(s) \exp(i\sigma_A(\Psi) \;t).
\label{Awaves1}
\end{equation}
The properties of perturbations with this form also occur in a magnetospheric context where they have been considered by 
\citet{Wright1999JGR}.
We have dropped the subscript $n$ on $\sigma_A(\Psi)$. The function $f_{A, \Psi}$  is the solution of  (\ref {CAW3}) for the corresponding continuum frequency $\sigma_A(\Psi)$ . In (\ref{Awaves1}) we have assumed that there is no phase difference between the standing continuum Alfv\'{e}n waves on different magnetic surfaces. Time $t$ in (\ref{Awaves1})  is dimensionless. It is equal to dimensional real time multiplied with  $\sigma_{A,R}$. The waves defined in (\ref{Awaves1}) are standing in the $\chi $ direction and (apparently) propagating in the $\Psi $ direction.  Let us now determine the apparent propagation of the phase in (\ref{Awaves1}). The motion defined in (\ref{Awaves1})  is multidimensional and its phase depends on position and on time.  Its dependence on position is in general not linear. For a multidimensional wave with phase $\varphi(\vec{x}, t)$ so that there is an exponential dependence
\begin{equation}
\exp(i \; \varphi(\vec{x},t)\;)
\label{GeneralPhase}
\end{equation}
 an instantaneous  local frequency $\sigma$ and an instantaneous  local wave vector $\vec{k}$ can be defined  as 
\begin{equation}
\sigma = \frac{\displaystyle \partial\varphi( \vec{x},t)} {\displaystyle \partial \;t}, \;\; \vec{k} = - \nabla \varphi( \vec{x},t).\;\;
\label{GFKV1}
\end{equation}
Note that the wave vector $\vec{k}$ is dimensionless. It is equal to the  dimensional physical wave vector  multiplied with 
$1/L_R$. A phase velocity can be defined in any direction \citep[see e.g.][]{BornWolf1999}. We adopt the traditional version and choose the direction normal to the wave front i.e. the direction of $\vec{k}$ so that
\begin{equation}
\vec{v}_{ph} = \frac{\displaystyle \sigma} {\displaystyle \mid \vec{k} \mid} \vec{1}_{k} .
\label{GPV1}
\end{equation}
Here $\varphi(\vec{x},t) = \sigma_A(\Psi) \; t$ so that the local frequency  $\sigma $ defined in (\ref{GFKV1})  is  $\sigma_A(\Psi)$. The local wave number $\vec{k}$ is
\begin{eqnarray}
\vec{k} & = & - t \;\nabla \sigma_A(\Psi) \nonumber \\
& = & -t \;\frac{\displaystyle d \sigma(\Psi)}{\displaystyle d \Psi} \;\nabla \Psi \nonumber \\
& = & - t \; \frac{\displaystyle d \sigma(\Psi)}{\displaystyle d \Psi} \; \mid \nabla \Psi \mid \vec{1}_{\Psi} \nonumber \\
& = &  - t \; \frac{\displaystyle d \sigma(\Psi)}{\displaystyle d \Psi} \; B(\Psi, \chi) \; \vec{1}_{\Psi} .
\label{PhaseVector}
\end{eqnarray}
The last line of (\ref{PhaseVector}) follows from the fact that 
$\mid \nabla \Psi \mid \; = \; \mid \vec{B}_p( \Psi, \chi) \mid \; =\;  B ( \Psi, \chi)$.
Equation  (\ref{PhaseVector}) tells us that the phase vector $\vec{k}$ is
antiparallel to $\nabla \sigma_A(\Psi)$. Hence   an increase /decrease  in $\sigma_A(\Psi)$ with $\Psi$ corresponds to apparent downward/upward propagation 
\begin{eqnarray}
\mbox{apparent upward propagation} & :& \;\; \frac{\displaystyle d \sigma_A(\Psi)}{\displaystyle d \Psi} < 0 \nonumber \\
\mbox{apparent downward propagation} & :& \;\; \frac{\displaystyle d \sigma_A(\Psi)}{\displaystyle d \Psi} > 0 
\label{UpDownAW}
\end{eqnarray}
Equation  (\ref{PhaseVector}) also shows that  (1) $\mid \vec{k} \mid$
increases linearly in time generating scales that decrease inversely proportional to time $t$ (as was also found by \citet{Mann1995JGR}),  
(2) $\vec{k}$ has only a component in the $\Psi$-direction i.e. normal to the magnetic surfaces and (3) $\vec{1}_k = \pm \vec{1}_{\Psi}$ where the $\pm$ sign corresponds to $d \sigma_A(\Psi) /  d \Psi <0, >0$.
The phase velocity $\vec{v}_{ph}$ is 
\begin{equation}
\vec{v}_{ph} = - \frac{\displaystyle 1}{\displaystyle t} \;
\frac{\displaystyle \sigma_A(\Psi)}{ \frac{\displaystyle d \sigma_A(\Psi)}{\displaystyle d \Psi}}\;
\frac{\displaystyle 1}{\displaystyle  B(\Psi, \chi) } \; \vec{1}_{\Psi} .
\label{PhSpeedAW1}
\end{equation}
Equation (\ref{PhSpeedAW1}) is a key result as it shows that there is an apparent propagation of phase when continuum Alfv\'{e}n waves are excited on magnetic surfaces. There is apparent upward/downward  propagation when $v_{ph}  > 0, \;\; v_{ph}  < 0$  which according to (\ref{PhSpeedAW1}) happens when (\ref{UpDownAW}) applies. In general $\vec{k}$ and $\vec{v}_{ph}$ are functions of $\Psi, \chi$ and time $t$.
In case the magnetic field strength and  the equilibrium density are flux functions we can rewrite equation 
(\ref{PhSpeedAW1}) as
\begin{equation}
\vec{v}_{ph} =  - \frac{\displaystyle 1}{\displaystyle t} \;
\frac{\displaystyle 1}{\displaystyle
\frac{\displaystyle 1}{\displaystyle v_A(\Psi)}   \frac{\displaystyle d v_A(\Psi)}{\displaystyle d \Psi}  -
 \frac{\displaystyle 1}{\displaystyle L(\Psi)}   \frac{\displaystyle d L(\Psi)}{\displaystyle d \Psi}} \;
 \frac{\displaystyle 1}{\displaystyle  B(\Psi) } \;\vec{1}_{\Psi} .
 \label{PhSpeedAWSpCase}
\end{equation}
The condition for  apparent upward/downward  propagation (\ref{UpDownAW}) can now be rewritten as 
\begin{eqnarray}
\mbox{upward propagation} & : & \;\; 
\frac{\displaystyle 1}{\displaystyle v_A(\Psi)}   \frac{\displaystyle d v_A(\Psi)}{\displaystyle d \Psi} <
\frac{\displaystyle 1}{\displaystyle L(\Psi)}   \frac{\displaystyle d L(\Psi)}{\displaystyle d \Psi} 
\nonumber \\
\mbox{downward propagation} & : & \;\; 
\frac{\displaystyle 1}{\displaystyle v_A(\Psi)}   \frac{\displaystyle d v_A(\Psi)}{\displaystyle d \Psi} >
\frac{\displaystyle 1}{\displaystyle L(\Psi)}   \frac{\displaystyle d L(\Psi)}{\displaystyle d \Psi}
\label{UpDownAWSpCase}
\end{eqnarray}
Equivalent formulae to our equations (\ref{UpDownAW})-(\ref{UpDownAWSpCase}) have been derived elsewhere and 
used to infer magnetospheric structure based upon the direction and speed of observed phase motion.
The interested reader is directed to the summary given by \citet{wright2006global}.
Regarding the closed magnetic field, the number of nodes can affect the phase velocity.
When one particular node is dominant, Eq. (\ref{PhSpeedAWSpCase}) and the condition (\ref{UpDownAWSpCase}) are available without any modification (substitute Eq.  (\ref{eq:periodiccont}) or 
(\ref{eq:periodiccont_s}) to Eq. (\ref{PhSpeedAW1})).  
If the multiple nodes exist on a magnetic surface, the appearance of the apparent propagation 
will become more complicated.
It is straightforward to adapt the preceding analysis to include an initial phase difference between the excited  Alfv\'{e}n continuum waves. We denote the initial phase difference as $\varphi_0(\Psi)$. The expression (\ref{PhSpeedAW1}) for the apparent propagation speed becomes
\begin{equation}
\vec{v}_{ph} =
 -\frac{\displaystyle \sigma_A(\Psi)}{  t\frac{\displaystyle  \;d \sigma_A(\Psi)}{\displaystyle d \Psi}    
+ \frac{\displaystyle \;d \varphi_0 (\Psi)}{\displaystyle d \Psi}} \;
\frac{\displaystyle 1}{\displaystyle  B( \Psi, \chi) } \; \vec{1}_{\Psi} .
\label{PhSpeedAW2}
\end{equation}

\begin{figure}[!h] 
\center{\includegraphics[bb=0 0 1133 1700,width=12cm]{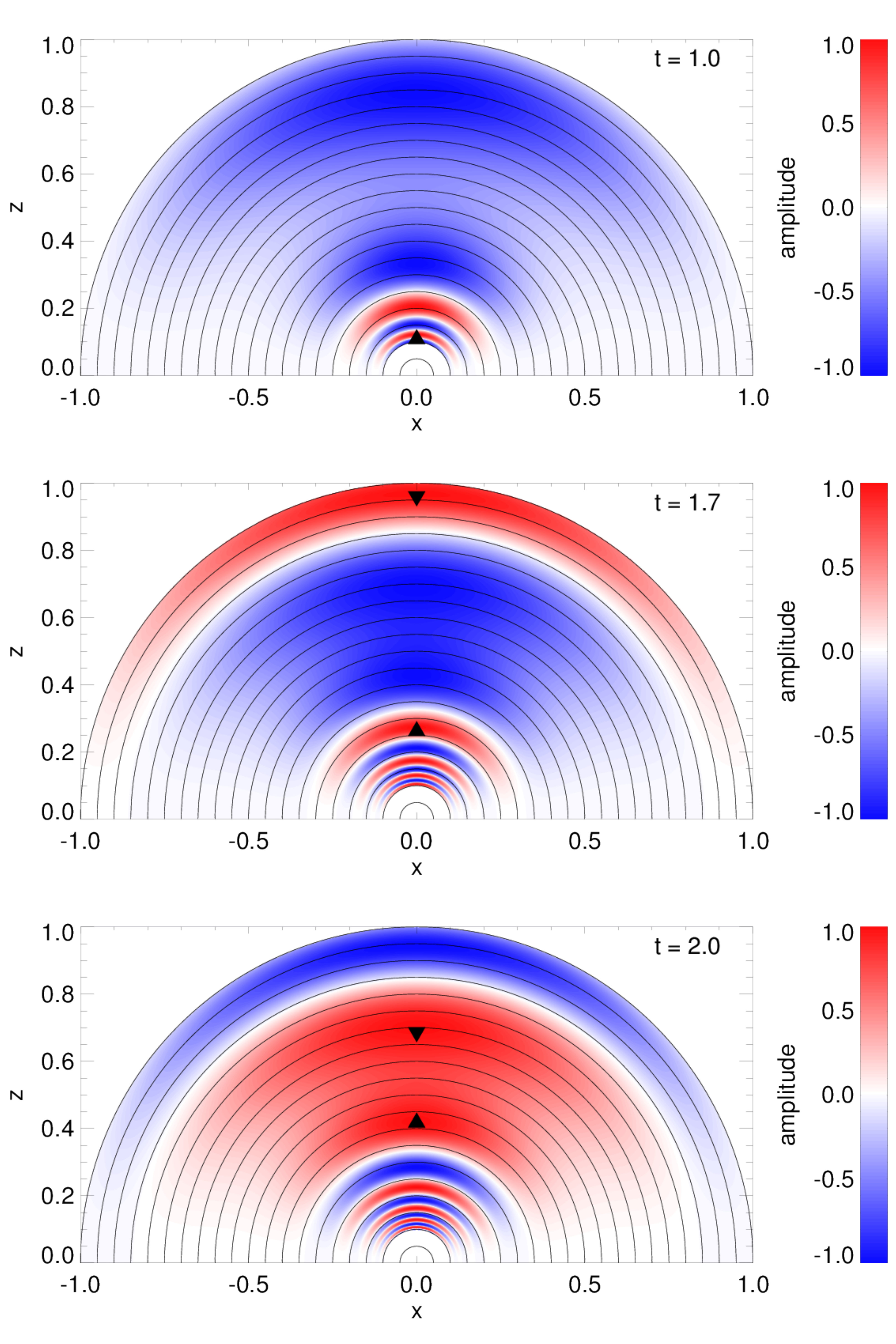}} \caption{\small
Snapshots of time evolution of Eq. (\ref{SimpleSol}). Color contour represents the amplitude orthogonal 
to the plane.
Solid lines show the assumed magnetic field of Eq. (\ref{CircleMag}). 
The triangles and inverse triangles mark the same phase ($\Phi = 20.5$), and show
the upward and downward propagation, respectively.}
\label{snapshots}
\end{figure}

\begin{figure}[!h] 
  \center{\includegraphics[bb=0 0 850 566,width=12cm]{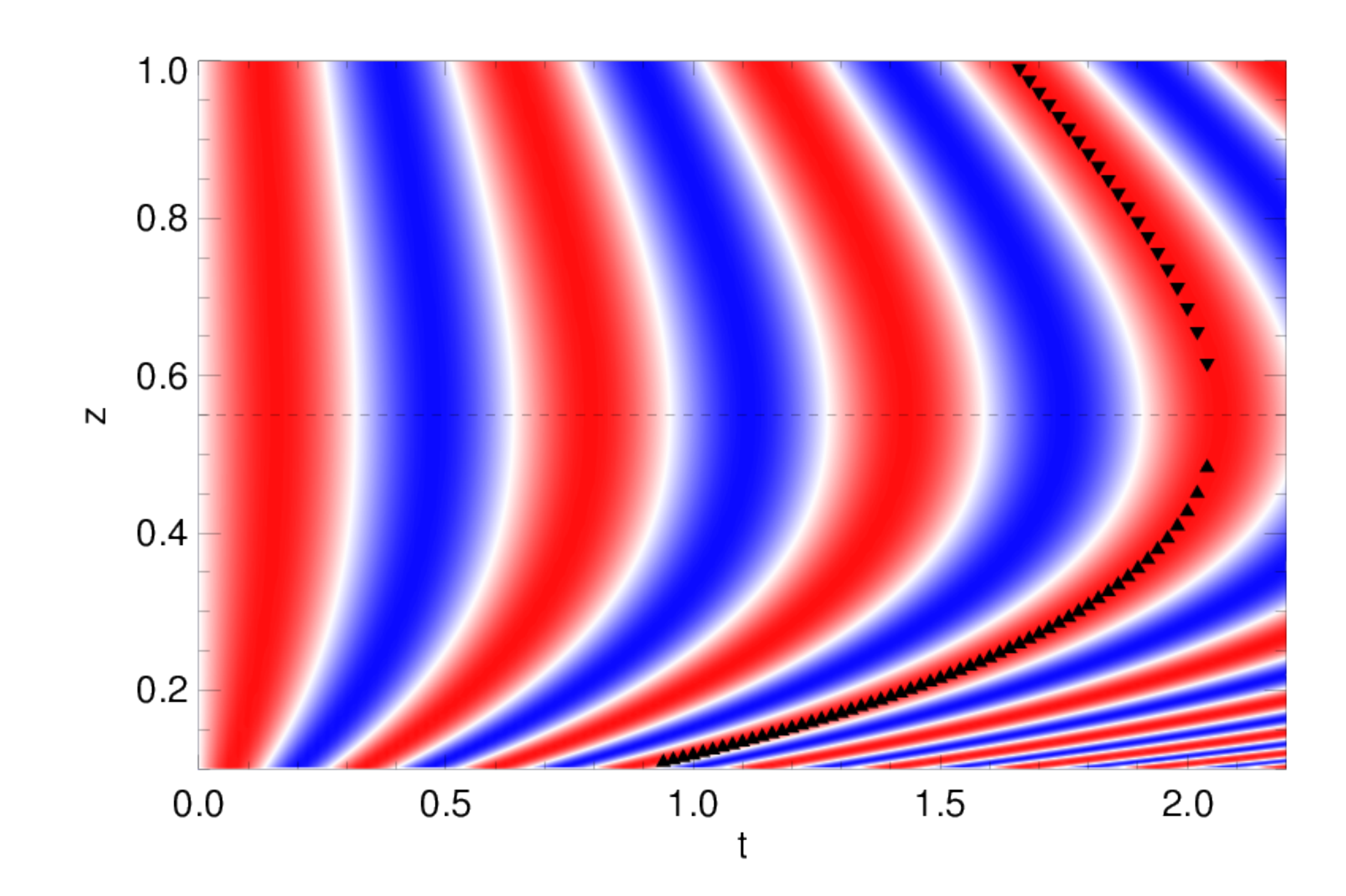}}  \caption{\small
Time-height plot along $x=0$ of Fig. \ref{snapshots}. Horizontal and vertical axis represents 
time $t$ and height $z$, respectively. Dashed line shows the border of upward and downward propagation
derived by criteria (\ref{UpDownCir}).
Color contour, triangles and inverse triangles represent the same meaning in Fig, \ref{snapshots}.}
\label{timeheight}
\end{figure}

%%%%%%%%%%%%%%%%%%%%%%%%%%%%%
\vspace{5mm}
Here, we present an example of upward and downward propagation by using a simple model.
We consider the semicircular magnetic field described as
\begin{equation}
  \vec{B}=\vec{1}_{\theta}~~(0<\theta <\pi), 
  \label{CircleMag}
\end{equation}
where $\vec{1}_{\theta}$ is the unit vector in the $\theta $ direction in a system of polar
coordinate in $(x,z)$ plane (see also the magnetic configuration in Fig. \ref{snapshots}).
In this particular case, the flux surface $\Psi =\Psi _{0}$ corresponds to $r=r_{0}$
due to the uniform unit field strength in the whole domain.  
The solution of the fundamental standing Alfv\'en wave on each flux surface $r$ is
\begin{equation}
  \xi _{y} (r,\theta ,t)=\sin \left( \sigma _{A}(r)t \right) \sin \theta , 
  \label{SimpleSol}
\end{equation}
where $\sigma _{A}(r)=2\pi v_{A} (r)/L(r)$ and $L(r)=\pi r$. 
Since the magnetic field is not force-free, a pressure gradient in the radial direction is required for magneto-static equilibrium. We do not concern here about pressure. Our focus is on continuum Alfv\'{e}n waves and from the discussion in subsection 3.1 we recall that pressure has not any effect on the Alfv\'{e}n continuum waves.
The distribution of Alfv\'en velocity is set as
\begin{equation}
  v_{A}(r) = \exp \left( \frac{r}{a} \right)
\end{equation}
where $a=0.55$.
In this setting,
the wave vector and phase speed of apparent cross-field propagation 
are computed  from Eqs. (\ref{PhaseVector}) and (\ref{PhSpeedAWSpCase}) as
\begin{eqnarray}
  \vec{k}&=&-\frac{2t}{r}\left( \frac{1}{a}-\frac{1}{r}\right) \exp \left( \frac{r}{a} \right)  \vec{1}_{r} ,
  \label{PhaseVectorCir} \\
  \vec{v}_{ph}&=&-\frac{1}{t}\frac{ar}{r-a}\vec{1}_{r} .
  \label{PhaseSpeedCir}
\end{eqnarray}
The condition for  upward/downward  propagation is 
\begin{eqnarray}
\mbox{upward propagation} & : & \;\; 
\frac{\displaystyle 1}{\displaystyle a}  < \frac{\displaystyle 1}{\displaystyle r}
\nonumber \\
\mbox{downward propagation} & : & \;\; 
\frac{\displaystyle 1}{\displaystyle a}  > \frac{\displaystyle 1}{\displaystyle r}
\label{UpDownCir}
\end{eqnarray}
Hence, the condition for upward propagation is satisfied for the region with $r<a$ and
that for downward propagation is satisfied in the region wit $r>a$.
Figure. \ref{snapshots} shows the snapshots of the time evolution of the standing wave solution 
described as  Eq. (\ref{SimpleSol}).
Figure. \ref{timeheight} is the time-height plot along $x=0$ of Fig. \ref{snapshots}.
The dividing line between the regions of  upward and downward propagation matches the criteria (\ref{UpDownCir}).
It is also evident in Figs. \ref{snapshots} and  \ref{timeheight} 
that the apparent wave vector increases with time as derived in Eq. (\ref{PhaseVectorCir}),  
which is the nature of phase mixing.
Moreover, the apparent phase speed is getting slower and slower as time goes on,
which agrees with Eq. (\ref{PhaseSpeedCir}). 
If the standing waves are retained for a long time before they are dissipated by some instabilities
or the local diffusivity during phase mixing, 
the phase speed is getting slower and slower, resulting in superslow propagation.

\vspace{5mm}

Let us now go back to the coronal arcade model of \citet{Oliver1993A&A}.
According to (\ref{PhaseVector}) and (\ref{UpDownAW}) it follows that the
apparent propagation of the phase is always upward for models with $\delta <3$.
For models with $\delta \geq 3$  the apparent propagation of phase is upward for
$\forall x_0 \in [0 \; ;\;x_C[$ and downward for $\forall x_0 \in ]x_C\;
;\;1]$. 

\begin{figure}[htbp]
\center{\includegraphics[bb=0 0 428 428,width=9cm]{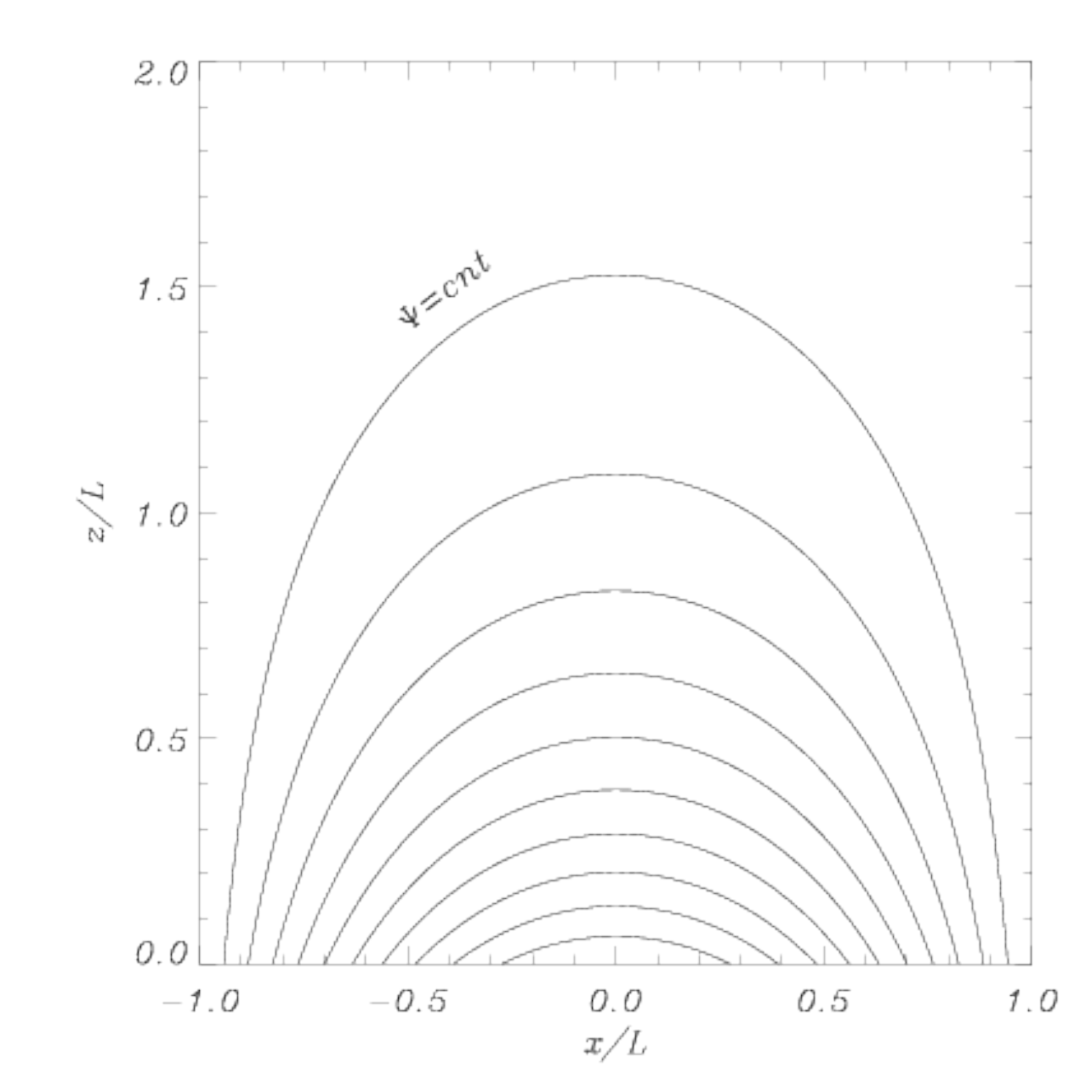}}
\caption{\small
Magnetic field lines in the potential coronal arcade studied by \citet{Oliver1993A&A}. The
magnetic field is given by $B_x=B_0 \cos\left(x/\Lambda_B\right) e^{-z/\Lambda_B}$, $B_z=-B_0
\sin\left(x/\Lambda_B\right) e^{-z/\Lambda_B}$, being $\Lambda_B=2L/\pi$ and $L$ half the arcade width. On each
field line $\Psi$ is constant. The magnetic field lines can be identified by their footpoints, 
$x_0$,  at $z=0$, or equivalently, by the value of $\Psi$.}\label{figurearcade}
\end{figure}

Let us focus on a particular example that might be illustrative of the application of
Equation (\ref{PhSpeedAW1}). \citet{Oliver1993A&A} found an analytic solution for the
potential arcade (see Figure \ref{figurearcade}) with $\delta=0$. In this case these authors
found that, for the fundamental mode with $n
=0$, the dependence of the Alfv\'en frequency on the footpoint position ($x_0$) is

\begin{equation}
\sigma_{A} =\frac{\pi}{2 x_0} v_{A0} \cos\left(x_0/\Lambda_B\right).
\label{sigmaA}
\end{equation}
In this expression $v_{A0}=B_0/\sqrt{\mu \rho_0}$ is the Alfv\'en speed at $z=0$ and is independent of the
$x-$coordinate.

Using the flux function it is straight forward to relate $\Psi$ with $x_0$, since
\begin{equation}
\Psi = -B_0 \Lambda_B\cos\left(x\right/\Lambda_B) e^{-z/\Lambda_B},
\label{psiflux}
\end{equation} 
at $z=0$ we simply have that
\begin{equation}
\Psi = -B_0 \Lambda_B \cos\left(x_0/\Lambda_B\right).
\label{psifluxx0}
\end{equation} 
The next step is to evaluate the derivative of the Alfv\'en frequency with respect to $\Psi$.
This is done using the fact that 
\begin{equation}
\frac{d \sigma_{A}}{d \Psi} = \frac{d \sigma_{A}}{d x_0} \frac{d x_0}{d \Psi}.
\label{eqdelta0}
\end{equation}
Using Equations (\ref{sigmaA}) and (\ref{psifluxx0}) we find
\begin{equation}
\frac{d \sigma_{A}}{d \Psi} =-\frac{\pi}{2x_0}v_{A0}\left[\frac{1}{x_0} \cos\left(
x_0/\Lambda_B\right)+\frac{1}{\Lambda_B}\sin\left(x_0/\Lambda_B\right)\right]\frac{1}{B_0 
\sin\left(
x_0/\Lambda_B\right)}.
\label{dsigmadpsi}
\end{equation}
Now we need the expression for  $B(\Psi, \chi)$. Since it depends on the coordinate along
the particular field line we concentrate at the position where it has a minimum and
therefore the phase velocity has a maximum. This position is simply at $x=0$ in our arcade
configuration, i.e., at the center of the arcade where there is only a horizontal component
of the magnetic field. The horizontal component of the field line crossing the
center of the arcade that has its footpoint at $x_0$ is $B_0 \cos\left(x_0/\Lambda_B\right)$.
Hence the phase velocity in the
$z-$direction at the center of the arcade  is
\begin{equation} 
v_{phz}= \frac{1}{t}\frac{\sin\left(
x_0/\Lambda_B\right)}{\frac{1}{x_0} \cos\left(
x_0/\Lambda_B\right)+\frac{1}{\Lambda_B} \sin\left(x_0/\Lambda_B\right)}.
\label{vpharcade}
\end{equation}
It is interesting to note that according to (\ref{vpharcade}) the apparent phase speed,
which is always pointing upwards for $\delta=0$, is
independent of the value of the magnetic field and only depends on the geometrical aspects
of the magnetic configuration.

Equation (\ref{vpharcade}) can be written in terms of height at the center of the arcade, at
$x=0$, by using the relationship with the footpoint position
\begin{equation}
e^{-z/\Lambda_B} = \cos\left(x_0/\Lambda_B\right).
\label{z_x0}
\end{equation}

\begin{figure}[htbp] 
\center{\includegraphics[bb=0 0 428 285,width=9cm]{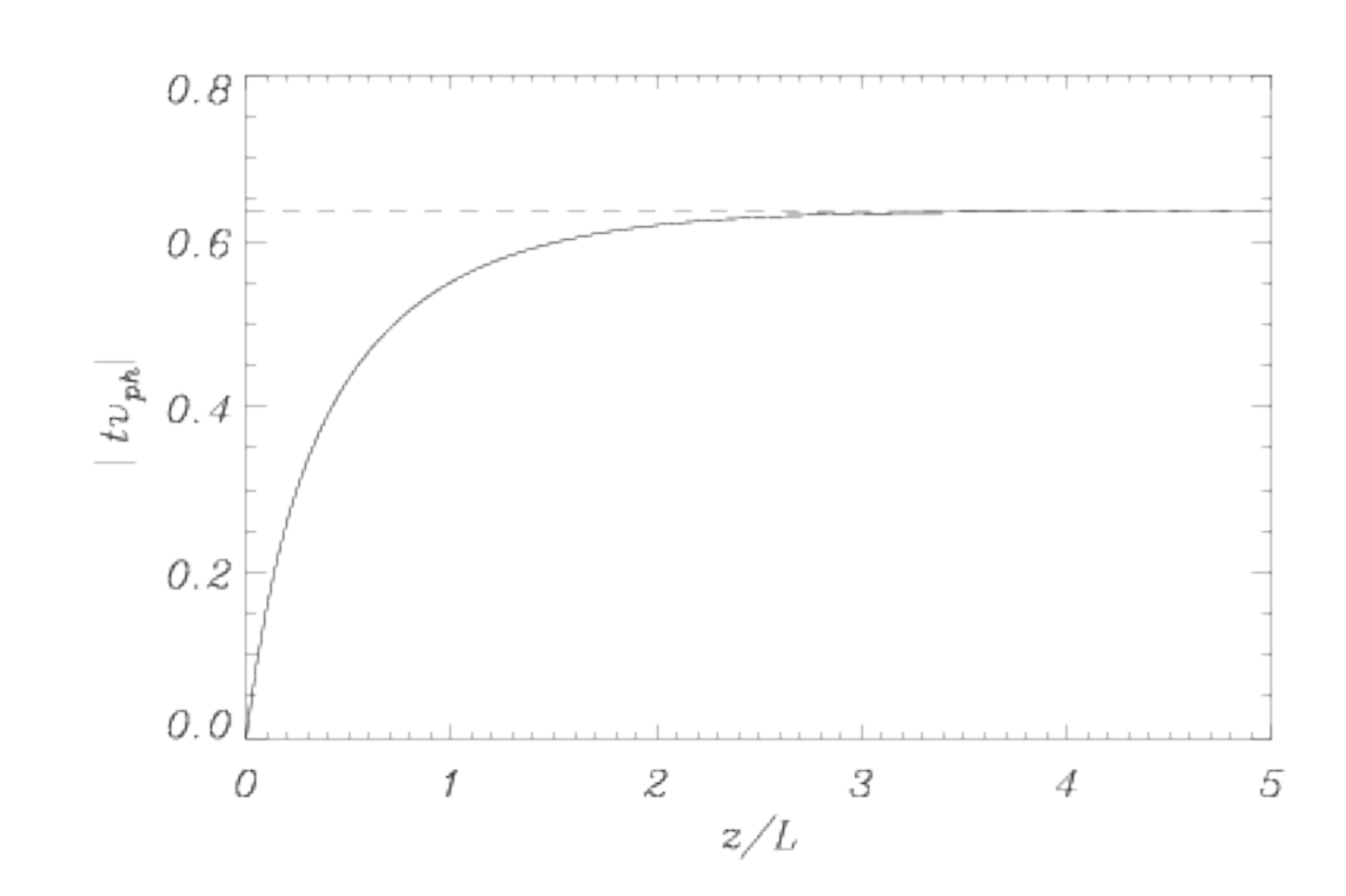}}
\caption{\small Modulus of apparent phase velocity in the $z-$direction of Alfv\'en waves as a
function of height at the center of the potential coronal arcade ($\delta=0$). The dashed line
corresponds to the limiting case $|tv_{ph}|=\Lambda_B$. For visualization purposes in this plot we have imposed that
$t=1$.}\label{figurevpharcade}
\end{figure}

We have used this expression to plot the dependence of the phase velocity as a function of
height in Figure \ref{figurevpharcade}. The absolute value of this magnitude decreases with
$z$ for low heights, i.e., small $x_0$. The dependence in this regime is of the form $
x_0^2/(t\Lambda_B)$. For large heights it asymptotically approaches to $\Lambda_B/t$ as it is inferred
from (\ref{dsigmadpsi}) in the limit of $x_0$ large. Hence the strongest change in the phase
speed takes place at low heights.

Now we turn to slow continuum modes and follow an equivalent process to that for Alfv\'en continuum modes. Let us consider a situation where the slow  continuum waves each with their own continuum frequency are excited on magnetic surfaces 
$\Psi_B \leq \Psi \leq \Psi_E$ with amplitude $S(\Psi)$   so that
\begin{equation}
\xi_{\chi}(\Psi, s,t) = S(\Psi) f_{S, \Psi}(s) \exp(i\sigma_S(\Psi) \;t) .
\label{Swaves1}
\end{equation}
The function $f_{S, \Psi}$  is the solution of  (\ref {CSW3}) for the corresponding continuum frequency $\sigma_S(\Psi)$. Since equation (\ref{Swaves1}) for continuum slow waves is formally identical to equation (\ref{Awaves1}) for continuum Alfv\'{e}n waves we can copy the equations for Alfv\'{e}n  continuum waves and replace 
$\sigma_A(\Psi) $  with $\sigma_S(\Psi) $.  In particular we can use equation  (\ref {PhSpeedAW2}) to 
find that the apparent propagation speed of the phase of slow continuum waves is 
\begin{equation}
v_{ph} = - \frac{\displaystyle \sigma_C(\Psi)}{  t\frac{\displaystyle  \;d \sigma_C(\Psi)}{\displaystyle d \Psi}    
+ \frac{\displaystyle \;d \varphi_0\Psi)}{\displaystyle d \Psi}}
\frac{\displaystyle 1}{\displaystyle  \mid \nabla \Psi \mid } \vec{1}_{\Psi} .
\label{PhSpeedSW12}
\end{equation}

\section{Application to Simulation Results}
\begin{figure}[htbp]
  \begin{center}
    \includegraphics[bb=0 0 283 226]{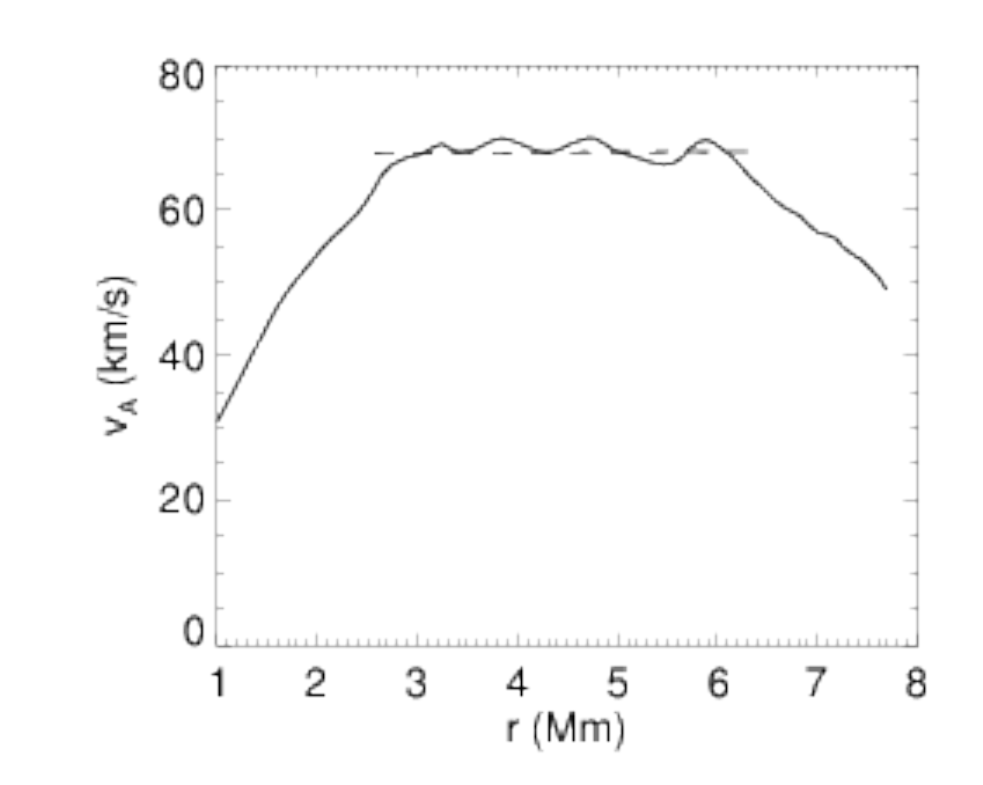}
    \caption{Mean Alfv\'{e}n velocities along the slit. The horizontal axis shows the distance along the slit.
    $r=0$ is the center of the flux rope.}
    \label{Alfven_profile}
 \end{center}
\end{figure}

\begin{figure}[htbp]
  \begin{center}
    \includegraphics[bb=0 0 283 226,scale=1.1]{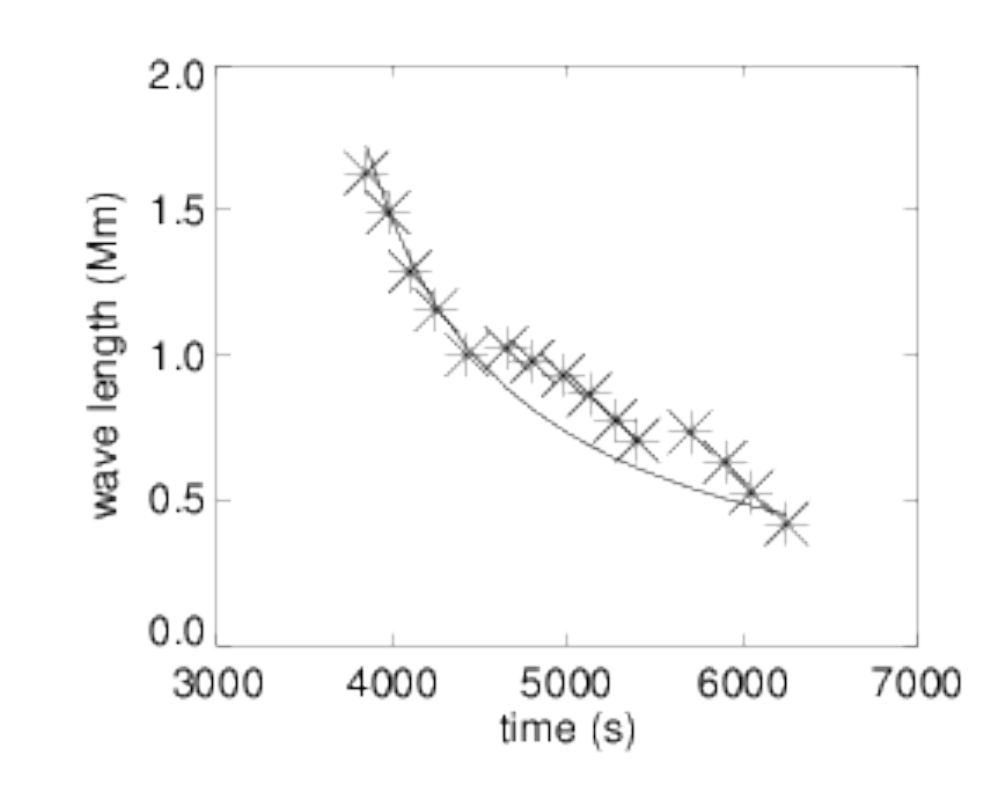}
    \caption{Wavelength in $r$-direction (dots) and the apparent wavelength 
      computed by Eq. (\ref{PhaseNum_KY}) (solid line) at $r=4$ Mm.}
    \label{wlength_4Mm}
  \end{center}
\end{figure}

In this section, we apply the theory of the apparent propagation due to 
the phase mixing to the cross-field superslow propagation 
in the simulation of \citet{KanekoYokoyama2015} shown as Fig. \ref{snapshots_los}. 
We show that this superslow cross-field propagation can be explained as caused by the continuum standing Alfv\'{e}n waves inside the flux rope.
Expressions for the apparent wave number and phase speed are given by Eqs.  (\ref{PhaseVector}) and  (\ref{PhSpeedAWSpCase}).
We regard the flux rope as a concentric cylinder, 
and assume $d\Phi /dr=B(r)$ and $\sigma _{A} (r) = 2\pi v_{A}(r)/L(r) = v_{A}(r)/r $
where $r$ is the distance along the slit and the length of one closed loop is $L(r)=2\pi r$.
We also assume that the Alfven waves inside the flux rope are excited simultaneously at time $t=t_{i}$.
Under these assumptions, the apparent wave number and phase speed
as a function of $r$ and $t$  are derived as 
\begin{equation}
  k(r,t) = -\frac{t-t_{i}}{r}\left( \displaystyle \frac{dv_{A}}{dr} - \frac{v_{A}}{r}\right), 
  \label{WaveNum_rt}
\end{equation}
\begin{equation}
  v_{ph}(r,t)=-\frac{1}{t-t_{i}}\frac{v_{A}(r)}{\displaystyle 
    \frac{dv_{A}(r)}{dr} - \frac{v_{A}(r)}{r}}.
  \label{PhaseVel_rt}
\end{equation}
Figure. \ref{Alfven_profile} shows
the profile of the mean Alfv\'{e}n velocity along the slit at time = $5000$ $\mrm{s}$.
We adopt the harmonic mean of Alfv\'{e}n velocities at the top and the bottom
of the loop as the representative value of the Alfv\'{e}n velocity on each magnetic surface.
Note that we use the in-plane Alfven velocity $v_{A}=\sqrt{(B_{x}^{2}+B_{z}^{2})/(4\pi \rho )}$
though the simulation includes the finite magnetic component perpendicular to the plane $B_{y}$.  
The reason for neglecting $B_{y}$ is that we are considering the projection of the wave path onto the plane (e.g. $L(r)=2\pi r$).
As shown in Fig. \ref{Alfven_profile},   
the mean Alfv\'{e}n velocities are constant in the region of $2.5$ $\mrm{Mm} < r <6.2$ $\mrm{Mm} $. 
Since $dv_{A}/dr=0$, Eqs.  (\ref{WaveNum_rt}) and (\ref{PhaseVel_rt}) can be simplified to
\begin{equation}
  \lambda (r,t)=\frac{2\pi}{k(r,t)}=\frac{2\pi r^{2}}{(t-t_{i})v_{A}}, 
\label{PhaseNum_KY}
\end{equation}
\begin{equation}
  v_{ph}(r,t)=\frac{r}{t-t_{i}}, 
\label{PhaseVel_KY}
\end{equation}
where $\lambda (r,t)$ is the apparent wavelength.
The dashed lines in Fig. \ref{snapshots_los} (c) are the use of 
Eq. (\ref{PhaseVel_KY}) for different values of $v_{ph}$.
We set $t_{i}=3000$ $\mrm{s}$ corresponding to the time when the waves are excited by radiative condensation.
Equation (\ref{PhaseVel_KY}) well explains 
the superslow phase speed in the simulation. 
Figure. \ref{wlength_4Mm} shows the wavelength in $r$-direction at $r = 4$ $\mrm{Mm}$. 
The solid line in Fig. \ref{wlength_4Mm} represents the apparent wavelength computed 
by inverse of Eq. (\ref{PhaseNum_KY})
with the local Alfv\'{e}n velocity of $v_{A}=70$ km at $r=4$ Mm (see Fig. \ref{Alfven_profile}).
The wavelength of the superslow propagation can be explained by Eq. (\ref{PhaseNum_KY}).
Our conclusion is that the cross-field superslow propagation in the simulation
is an apparent phenomenon due to the phase mixing of continuum Alfv\'{e}n waves in the flux rope.
The formulae derived in the previous section 
can correctly predict the apparent wavelength and phase speed.

\section{Discussion}

In this paper we  explored the continuous MHD spectrum for 2D equilibrium models in Cartesian geometry that are invariant in the $y$-direction and have a purely poloidal magnetic field. The actual equilibrium configurations that we have in mind are 2D coronal arcades \citet[see, e.g][]{Oliver1993A&A}. We showed that continuum  Alfv\'{e}n waves and continuum  slow waves that live on individual magnetic surfaces are phase mixed as time evolves. This process creates the illusion of waves propagating across the magnetic surfaces at very slow velocities. This phenomenon could be erroneously interpreted as fast magnetosonic waves. We derived expressions for the apparent cross-field phase velocity. This quantity depends on the  spatial variation of the local Alfv\'{e}n/slow frequency across the magnetic surfaces. For typical conditions in solar coronal arcades, the apparent phase velocity is  slower than the local Alfv\'{e}n/sound velocities.

The theory developed in the present paper can be used to understand the numerical simulations of \citet{Rial2010ApJ,Rial2013ApJ} and \citet{KanekoYokoyama2015} of MHD waves in coronal arcades and prominences. These authors obtained in their simulations the apparent superslow propagation discussed here.  For instance, \citet{KanekoYokoyama2015} find isotropic propagation at a superslow speed of  $3 \pm 2$ km/s to be compared with  a fast wave velocity of about $160 $ km/s and a slow wave velocity of about $70 $ km/s, and the speed gets slower with time. 
The superslow isotropic propagation found in the simulation of \citet{KanekoYokoyama2015} can be explained as apparent propagation due to continuum waves. 
The phenomenon of apparent propagation should be taken into account in the future to correctly analyze the result of numerical simulations.

In addition, apparent propagation may be an alternative explanation of the recent observations by \citet{Schmieder2013ApJ} of MHD waves in a solar prominence with an essentially horizontal magnetic field.  \citet{Schmieder2013ApJ} report  upward propagation with a speed of $ 5 \pm 3$ km/s and downward propagation with a velocity of $ 5 \pm 2$ km/s. \citet{Schmieder2013ApJ} interpreted their observations with a model based on fast magnetosonic waves. However, the interpretation in terms of fast magnetosonic waves poses a problem  since the velocity of the observed propagation is  slower than that associated with a fast  wave.  In order to arrive at phase velocities comparable to the observed velocities \citet{Schmieder2013ApJ} had to assume values of the density larger than the typical prominence densities and relatively small projection angles. 
As can be seen from Eq. (\ref{PhSpeedAW1}) the speed of apparent propagation depends on time as $1/t$. This implies that it is very slow for large values of time $t$,  i.e. when the waves are observed long after their excitation, but fast right after the excitation of the waves. Also, the spatial variation of the local Alfv\'{e}n frequency $\sigma _{A}(\Psi )$ plays a role. A slow spatial variation of $\sigma _{A}(\Psi )$ causes a rapid propagation while a fast variation of $\sigma _{A}(\Psi )$ leads to slow propagation.
The observations of \citet{Schmieder2013ApJ} could  also be interpreted as apparent waves, which would naturally have an apparent phase velocity smaller than the  velocity associated to a fast wave. A detailed investigation of those observation is however beyond the scope of the present paper and is left for future works. 

The present paper offers, to the best of our knowledge, the first detailed discussion of superslow propagation due to continuum waves in solar physics. However, this physical mechanism is common to MHD waves in a variety of non-uniform plasmas environments. It has been brought to our attention by the referee that this phenomenon has been studied in the context of magnetospheric physics \citep[e.g.][]{Mann1995JGR,Wright1999JGR} and reviewed by \cite{wright2006global}. 
Actually, some formulae in section  4 correspond to the equations derived in \citet{Mann1995JGR} and \citet{Wright1999JGR}.
In magnetospheric physics  
the phenomenon of phase motion is an important element in the discussion to explain 
the generation of Alfv\'{e}n waves. This generation involves the Alfv\'{e}n resonance and requires
that the turning point of the fast wave is sufficiently close to the resonant point so that the wave has to
tunnel only over a short distance to get to the resonant point. In the solar case, it can be argued that
many slow and Alf\'{e}n waves are excited at the photosphere by convective motions and propagate to 
the chromosphere and corona. The solar atmosphere is a very likely place to find apparent propagation
due to phase mixing. The apparent motion can be a clue to find evidence for phase mixing in the solar 
atmosphere. 

\acknowledgments
We are grateful for the referee's insightful and fruirful suggestions. 
We are strongly encouraged and benefited by the referee's preliminary analyses. 
TK was supported by the Program for Leading Graduate School, MEXT, Japan.
This work was supported by JSPS KAKENHI Grant Number 15H03640.
Numerical computations were in part carried out on Cray XC30 at Center for Computational Astrophysics, National Observatory of Japan.
RS acknowledges support from MINECO through project AYA2014-54485-P and from FEDER funds. RS also acknowledges support from MINECO through a `Juan de la Cierva' grant, from MECD through project CEF11-0012, and from the `Vicerectorat d'Investigaci\'o i Postgrau' of the UIB.
JT acknowledges support from the Spanish Ministerio de
Educaci\'on y Ciencia through a Ram\'on y Cajal grant.
JT acknowledges support from MINECO through project AYA2014-54485-P 
and from FEDER funds.
MG was supported by IAP P7/08 CHARM (Belspo) and the GOA-2015-014 (KU Leuven).
TVD was supported by an Odysseus grant of the FWO Vlaanderen, the IAP P7/08
CHARM (Belspo) and the GOA-2015-014 (KU~Leuven).

\appendix

%% \bibliographystyle{apj}
%% \bibliography{reference}

\end{document}